\documentclass[twocolumn]{aastex62}
\pdfoutput=1

\usepackage[binary-units]{siunitx}
\usepackage[natbib]{}
\usepackage{graphicx}
\usepackage{bm}
\usepackage{mathtools}
\usepackage{siunitx}
\usepackage{multirow}

\graphicspath{{./}}

\shorttitle{Biases and Cosmic Variance in Molecular Gas Abundance Measurements}
\shortauthors{Keenan et al.}

\begin{document}

\title{Biases and Cosmic Variance in Molecular Gas Abundance Measurements at High Redshift}

\correspondingauthor{R. P. Keenan}
\author[0000-0003-1859-9640]{Ryan P. Keenan}
\altaffiliation{NSF Graduate Research Fellow}
\affiliation{Steward Observatory, University of Arizona, 933 North Cherry Avenue, Tucson, AZ 85721, USA}
\email{rpkeenan@email.arizona.edu}

\author[0000-0002-2367-1080]{Daniel P. Marrone}
\affiliation{Steward Observatory, University of Arizona, 933 North Cherry Avenue, Tucson, AZ 85721, USA}

\author[0000-0002-3490-146X]{Garrett K. Keating}
\affiliation{Harvard-Smithsonian Center for Astrophysics, 60 Garden Street, Cambridge, MA 02138, USA}

\begin{abstract}

Recent deep millimeter-wave surveys attempt to measure the carbon monoxide (CO) luminosity function and mean molecular gas density through blind detections of CO emission lines. While the cosmic star formation rate density is now constrained in fields hundreds of arcmin$^2$ or more, molecular gas studies have been limited to $\le50$ arcmin$^2$. These small fields result in significant biases that have not been accounted for in published results. To quantify these biases, we assign CO luminosities to halos in cosmological simulations to produce mock observations for a range of field sizes. We find that fields $\lesssim 10$ arcmin$^2$ alter the recovered shape of the luminosity function, causing underestimates of the number of bright objects. Our models suggest that current surveys are sensitive enough to detect sources responsible for approximately half of the cosmic molecular gas density at high redshift. However, uncertainties in the gas density measurement are large, and cosmic variance may double the uncertainty claimed in these surveys. As a result, the field size needed to detect redshift evolution in the molecular gas at high confidence may be more than an order of magnitude larger than what current surveys have achieved. Shot power intensity mapping measurements are particularly sensitive to Poisson variance and require yet larger areas to constrain the gas density or its evolution. We provide a simple prescription for approximating uncertainty in total CO emission as a function of survey area and redshift, for both direct detection and intensity mapping surveys.
\end{abstract}

\section{Introduction} \label{sec:intro}

Deep ultraviolet, optical, and infrared surveys have been used to constrain the evolution of the UV and IR luminosity functions and star formation rate density (SFRD) to redshifts $z>8$ \citep{madau+14,bouwens+15,finkelstein+15,driver+18}. These studies have found that the SFRD evolves significantly with redshift, peaking at $z\sim2$ then falling an order of magnitude to present day. It is expected that changes in the abundance and properties of molecular gas, the raw material for star formation, are the driver of the changing cosmic star formation rates \citep{tacconi+18}. The history of the molecular gas abundance is therefore a necessary component of our understanding of galaxy evolution \citep{carilli+13}. Over the past decade, advances in capabilities for millimeter astronomy have made it possible to conduct complementary, large surveys of molecular gas using emission from the $^{12}$CO molecule (hereafter CO). These projects aim to constrain redshift evolution of molecular gas density through measurements of the luminosity function of CO rotational emission lines, allowing for detailed comparisons with the SFRD history.

With results from the first generation of high redshift CO surveys now available \citep{decarli+14,decarli+16,keating+16,riechers+18,decarli+19,lenkic+20,keating+20arxiv}, it is important to carefully interpret the measured luminosity functions \citep{popping+19}. In particular, current surveys have been limited to areas significantly smaller than the optical and IR deep fields used in measuring SFRD. Rare objects at the bright end of the CO luminosity function, with large molecular gas reservoirs, can be important contributors to the molecular gas density \citep{carilli+16,lagos+15,popping+15}. Surveys covering only a few arcmin$^2$ may not sample enough volume to reliably recover such objects, which could bias many of their reported results. Accurate comparison to both observational data at other wavelengths and theoretical predictions using cosmological simulations and semi-analytical models will not be possible if these biases are not understood and accounted for.

At the same time, the number density of faint objects can vary considerably from field to field, depending on the large scale structures present in the region sampled. Numerous studies have found that this cosmic variance has a sizable effect on galaxy number counts and luminosity functions at optical wavelengths. \citet{moster+11} use simulations to conclude that uncertainty due to cosmic variance usually exceeds Poisson variance in the optical deep fields used for galaxy evolution studies. \citet{driver+10} come to a similar conclusion based on empirical measurements of cosmic variance in the local universe using SDSS. \citet{trenti+08} even suggest that cosmic variance can bias the shape determined by parametric fits of the luminosity function. 

These effects have not received great attention in the context of molecular gas. \citet{popping+19} considered sample variance in the limited context of the 5 arcmin$^2$ ASPECS study using semi-analytical models of the molecular gas mass function. Most existing observational studies have instead relied on optical results which are not necessarily well matched to the types of objects and surveys employed for CO, or have dismissed cosmic variance altogether. To clarify the role of sample variance effects in molecular gas surveys, a more complete exploration of the topic is warranted. 

In this paper we use cosmological simulations of large volumes to explore the biases and uncertainties present in studies of molecular gas abundance at high redshift. Simulations of millions of cubic megaparsecs with mass resolution better than $10^8$ M$_\odot$ now exist, allowing us to conduct mock observations using thousands of realizations of a field \citep[e.g.][]{TNG1,MDsim,eaglesim,illustrissim}. Using a model for assigning CO luminosity to dark matter halos, we construct catalogs of CO emitters and generate ensembles of light cones that sample different regions of the parent simulation volume. From these light cones we assess how well surveys of different sizes do at recovering the true values of quantities related to the CO luminosity function, and develop a prescription for quantifying cosmic variance optimized for molecular gas surveys.

The remainder of this paper is organized as follows: in Section~\ref{sec:observables} we describe the methods for surveying CO at high redshift and the quantities measured by each. In Section~\ref{sec:model} we describe our procedure for creating an ensemble of CO light cones for mock observations. We describe observational biases and uncertainties in measurements of the CO luminosity function in Section~\ref{sec:direct} and measurements of moments of the luminosity function in Section~\ref{sec:moments}. We then consider how these results affect efforts to detect the redshift evolution of molecular gas properties in Section~\ref{sec:rs}. In Section~\ref{sec:modeldependence} we verify that our results are independent of our choice of model parameters, although their degree varies depending on the underlying properties of the CO emitting objects. In Section~\ref{sec:discussion} we discuss our results in the context of ongoing efforts to measure the shape of the luminosity function (\ref{sec:directdiscussion}), the cosmic molecular gas density (\ref{sec:rhodiscussion}), the CO brightness fluctuation power spectrum (\ref{sec:imdiscussion}), and dust mass function (\ref{sec:dustdiscussion}). We present our main conclusions in Section~\ref{sec:conclusion}. We present a prescription for estimating cosmic variance in surveys at a range of redshifts and survey geometries in Appendix~\ref{sec:prescription}. Throughout this paper we assume a flat $\Lambda$CDM cosmology with $H_0=67.74$ and $\Omega_m=0.31$ \citep{Planck15}, chosen for consistency with the IllustrisTNG simulations from which we generate our light cones.

\section{Target Quantities and Observables}\label{sec:observables}

CO emission lines are the most commonly used proxy for molecular gas \citep{bolatto+13}. The line luminosity of the CO(1-0) rotational transition is translated to a molecular gas mass via 
\begin{equation}
    M_{\rm mol} = \alpha_{\rm CO} L^\prime_{\rm CO}\, ,
\end{equation}
where $M_{\rm mol}$ is the mass contained in the molecular gas phase, $L^\prime_{\rm CO}$ is the observed CO luminosity and $\alpha_{\rm CO}$ is the conversion factor between them. The conversion factor, including helium\footnote{Note that whether $M_{\rm mol}$ is defined to include the contribution of Helium to the total mass of gas in the molecular phase varies among different works. Helium accounts for $\sim36$\% of the molecular gas mass.}, is found to be around 4.6 M$_\odot$ (K km s$^{-1}$ pc$^2$)$^{-1}$ for the Milky Way and normal star forming galaxies at higher redshifts \citep{cassata+20,carleton+17,daddi+10}. Galaxies undergoing intense starbursts show lower values, $\alpha_{\rm CO}\sim1$ \citep{downes+98}. This factor also varies as a function of metallicity \citep{narayanan+12}.

The CO luminosity function $\phi_{\rm CO}(L^\prime_{\rm CO},z) = dn_{\rm CO}/dL^\prime_{\rm CO}$ describes the differential number density of sources of luminosity $L^\prime_{\rm CO}$ at redshift $z$ per unit volume (we will hereafter drop the CO subscripts on $L^\prime$ and $\phi$ and the explicit redshift dependence from our notation). The molecular gas density at redshift $z$ can be calculated
\begin{equation}\label{eq:rho}
    \rho_{\rm mol}(z) = \int_0^\infty \alpha_{\rm CO} L^\prime \phi(L^\prime) dL^\prime
\end{equation}
if $\alpha_{\rm CO}$ can be treated as approximately constant, at least for galaxies responsible for the majority of the above integral, this equation simplifies to
\begin{equation}\label{eq:rhoT}
    \rho_{\rm mol}(z) = \alpha_{\rm CO} \mu_1
\end{equation}
where $\mu_1 = \int L^\prime \phi(L^\prime) dL^\prime$ is the first moment of the luminosity function.

\subsection{Measurement Approaches}

A measurement of the cosmic molecular gas density can therefore be made by constraining of the CO luminosity function. A number of approaches to this measurement have been pursued. 

At low redshift where complete and well understood catalogs of galaxies are available, the CO luminosity function can be constructed through targeted observations of a large sample. These surveys must have a simple selection function (e.g. all galaxies above a stellar mass threshold). Each target can then be weighted according to the fraction of the total selection parameter space it represents. The Extended CO Legacy Database for the GALEX Arecibo SDSS Survey (xCOLD GASS), measured the CO luminosity function at $z\sim0$ through targeted observations \citep{saintonge+17} of 532 local galaxies selected in bins of stellar mass from $10^9$ M$_\odot$ to $>10^{11.5}$ M$_\odot$. This targeted approach would be challenging at high redshift with the sensitivity of current instruments.

At high redshift two approaches have been employed. The first, which we will refer to as ``direct measurement,'' entails conducting blind, integral field spectroscopic surveys of a selected volume and searching for CO emission lines. These surveys search for CO emission lines by using large interferometers to scan a wide frequency band. Single-line detections are generally not enough to uniquely determine a redshift, as multiple CO transitions or other lines can redshift to the same frequency. Cross-matching with optical and near IR catalogs can allow for redshift determination and identification of the lines \citep{boogaard+19}. Once emission lines have been identified, their CO luminosities can be determined from their measured redshifts and fluxes, and the luminosity function can be determined directly by counting sources in bins of luminosity \citep{decarli+16,riechers+18,decarli+19}.

It is common practice to fit a parameterized form of the luminosity function to observed galaxy counts. The form most often assumed in the CO literature is a Schechter function \citep{schechter76}:
\begin{equation}\label{eq:schechter}
    \phi(L^\prime)dL^\prime = \phi_* \Big(\frac{L^\prime}{L_*}\Big)^\alpha \exp{\Big(-\frac{L^\prime}{L_*}\Big)} \frac{dL^\prime}{L_*}\, ,
\end{equation}
where $\phi_*$, $L_*$, and $\alpha$ are fit parameters describing the normalization, turnover between polynomial and exponential shape, and the slope of the polynomial portion of the function. Note that numerous parameterizations of this function exist with differing coefficients. We use the form in Equation~\ref{eq:schechter} here, and summarize the other forms in Appendix~\ref{appendix:schechter}. 

A second approach, called line intensity mapping, involves conducting spectroscopic surveys of larger volumes at lower sensitivity, and using intensity fluctuations in the resultant data cubes to statistically measure moments of the CO luminosity function without needing detections of individual galaxies \citep{visbal+10,lidz+11,gong+11,breysse+14,li+16}. This type of measurement can be done using single dish telescopes with multi-pixel receivers, making it possible to map large areas at much lower cost than via the direct survey approach. 

The primary observed quantity in intensity mapping observations is the power spectrum, which describes the contribution of intensity fluctuations on different scales to the total power in the map. It can be parameterized as 
\begin{equation}\label{eq:ps}
    P_{\rm CO}(k) = P_{\rm lin}(k) b_{\rm CO}^2 \mu_1^2 + \mu_2\,,
\end{equation}
where $P_{\rm CO}(k)$ measures the magnitude of fluctuations of spatial wavenumber $k$ in the CO intensity map, $P_{\rm lin}(k)$ is the underlying matter density power spectrum, $b_{\rm CO}$ is the tracer bias for CO emitters, and $\mu_2$ is the second moment of the luminosity function, $\mu_2 = \int_0^\infty L^{\prime 2}\phi(L^\prime)dL^\prime\,.$

The first term on the right-hand side of Equation~\ref{eq:ps} is referred to as the clustering power, and is proportional to the first moment of the CO luminosity function. The second term is frequently referred to in intensity mapping as the shot power.

Owing to the shape of the matter power spectrum, at large spatial scales (small $k$) the clustering power term dominates, typically by multiple orders of magnitude, while at small scales (large $k$) the shot power becomes similarly dominant. Thus intensity maps covering large areas with adequate spatial and spectral resolution can constrain both terms.

If the matter power spectrum and tracer bias can be estimated by other means \citep[e.g.][]{barkana+05}, the clustering power can be used to determine $\mu_1$. We can then estimate $\rho_{\rm mol}$ using Equation~\ref{eq:rhoT}. A parameterized version of the full luminosity function can also be inferred by jointly fitting the two moments. Degeneracies between parameters in the luminosity function mean that additional information is required for an optimal fit. This may come from the intensity mapping survey itself, which can constrain the bright end of the luminosity function through direct detections or upper limits on the brightest galaxies \citep{keating+16}. It may also be derived from the bright end measurements from direct detection surveys. This situation is in principle, no worse than for current direct detection efforts, where most surveys lack the dynamic range in luminosity required to constrain all parameters of the luminosity function.

Line intensity mapping is well suited to surveys over larger areas than direct measurements. On the other hand, moments of the luminosity function are weighted integrals that up-weight bright, rare galaxies and therefore may be more susceptible to bias and Poisson variance than the direct approach. We investigate these effects in detail in subsequent sections.

\subsection{Units of the Measured Quantities}

Following the convention used in most papers dealing with the CO luminosity function, we present luminosities in observer's luminosity (frequently denoted $L^\prime$) units of K km s$^{-1}$ pc$^2$. This can be related to solar luminosity units as \citep{carilli+13}:
\begin{equation}\label{eq:lp}
    L[L_\odot] = 3\times10^{-11} \nu_{\rm rest}^3 L^\prime[{\rm K\ km\ s^{-1}\ pc^2}]\,,
\end{equation}
where $\nu_{\rm rest}$ is the rest frequency of the emission line in GHz, 115.27 GHz for CO(1-0).

For moments of the luminosity function, we convert luminosity to units of $\mu$K Mpc$^{3}$ by multiplying by the conversion factor between radial velocity and luminosity distance
\begin{equation}\label{eq:drdv}
    \frac{dr}{dv}(z) = \frac{(1+z)^2}{H(z)}\,,
\end{equation}
where $z$ is the central redshift of the observations and $H(z)$ is the Hubble parameter at that redshift. This puts the first and second moments in units of $\mu$K and $\mu$K$^2$~Mpc$^3$ respectively. In these units, the first moment is referred to as the mean brightness temperature and denoted by $\langle T \rangle$ and the second moment is referred to as the shot power and denoted by $P_{\rm shot}$.

Note that since $dr/dv$ depends on redshift, constant mean brightness temperature or shot power with redshift does not mean that the corresponding physical quantities are not evolving. We can combine Equations~\ref{eq:rhoT} and~\ref{eq:drdv} to write the mean molecular gas density in terms of mean brightness temperature as
\begin{equation}\label{eq:Trho}
    \rho_{\rm mol} = \alpha_{\rm CO} \frac{H(z)}{(1+z)^2} \langle T \rangle\,.
\end{equation}

\subsection{Sources of Variance}
If galaxies are randomly distributed throughout the universe, with mean number density $n$, then the number of objects, $N_{\rm obs}$, appearing in a survey covering a small portion of the sky with volume $V_{\rm obs}$ is well approximated by a Poisson distribution \citep{kelly+08}. The mean (and variance) on $N$ will be $\langle N\rangle = n V_{\rm obs}$. We will refer to the variance in survey results due to such processes as Poisson variance.

The Poisson variance in the luminosity function around luminosity $L^\prime$ in a bin of size $\Delta L^\prime$ is then given by
\begin{equation}\label{eq:pvlf}
    \sigma_{\phi,\rm pois}^2 = \frac{1}{V^2\Delta L^{\prime2}} \langle N_{L^\prime}\rangle = \frac{1}{V\Delta L^\prime} \phi(L^\prime)\,,
\end{equation}
where $\langle N_{L^\prime}\rangle = V \phi(L^\prime) \Delta L^\prime$ is the mean number of galaxies in the bin. The Poisson variance for the $m$th moment of the luminosity function $\mu_m$ is given by multiplying the above result by $L^{\prime 2m} \Delta L^\prime$ and summing over all bins:
\begin{equation}\label{eq:pvm}
     \sigma^2_{\mu_m,{\rm pois}} = \frac{1}{V} \int_0^\infty L^{\prime 2m} \phi(L^\prime) dL^\prime\,.
\end{equation}

In reality, objects are not randomly distributed. When sampling the universe over volumes much larger than the largest coherent structures, sampling uncertainty on number counts of galaxies should be dominated by Poisson errors (which should in turn be fractionally small because of the large number of objects included in such a volume). However, when the volume probed is comparable to or smaller than the scale of large scale structure, a given sample will contain over- (under-) abundances of galaxies because it happens to sample over- (under-) densities in the underlying dark matter field. This increased variability is referred to as cosmic variance. High redshift surveys can be highly susceptible to cosmic variance because achieving the necessary depth for high redshift studies typically limits them to small fields.

The magnitude of cosmic variance can be expressed analytically and computed using linear theory \citep{moster+11}. However, nonlinear effects can increase the level of variance over results from linear theory alone \citep{munoz+10}. Cosmological simulations can be used to capture the full effect of large scale structure on the luminosity function and its moments. Therefore, in this paper we calculate the cosmic variance as
\begin{equation}\label{eq:cv}
    \sigma_{\rm cosmic}^2 = \sigma^2 - \sigma_{\rm pois}^2\,,
\end{equation}
where $\sigma^2$ is the total sample variance, which we measure directly from our simulated observations and $\sigma_{\rm pois}^2$ is the Poisson variance computed using Equations~\ref{eq:pvlf} and~\ref{eq:pvm} or the Poisson variance only simulations described in Section~\ref{sec:model}.

\subsection{Summary of Existing Measurements}
\begin{deluxetable*}{cccc|ccccc}
\tablecaption{Parameters of Existing Direct Detection and Intensity Mapping Surveys \label{tab:surveys}}
\tablehead{
    \colhead{} & \colhead{Number} & \colhead{Total} & \colhead{Frequency} & \multicolumn{5}{c}{Redshift Coverage by CO Transition} \\
    \colhead{Survey} & \colhead{of Fields} & \colhead{Area (am$^2$)} & \colhead{Range (GHz)} & \colhead{CO(1-0)} & \colhead{CO(2-1)} & \colhead{CO(3-2)} & \colhead{CO(4-3)} & \colhead{CO(5-4)}
    }
\startdata
    PdBI & 1 & 0.7 & 79--115 & & 1.0--1.9 & 2.0--3.3 & 3.0--4.8 & 4.0--6.2 \\
    ASPECS Pilot & 1 & 0.9 & 84--115 & & 1.0--1.7 & 2.0--3.1 & 3.0--4.5 & 4.0--5.9 \\
    & & & 212--272 & & & 0.3--0.6 & 0.7--1.2 & \\
    ASPECS LP & 1 & 4.6 & 84--115 & & 1.0--1.7 & 2.0--3.1 & 3.0--4.5 & 4.0--5.9 \\
    COLDz & 2 & $51 + 9$ & 30--39$^a$ & 2.0--2.8 & 4.9--6.7 & & & \\
    PHIBSS2 & 110 & 130 & Varies$^b$ & & 0.0--1.6 & 0.5--2.8 & 1.0--4.1 & 1.5--5.4 \\
    \hline
    COPSS I & 44 & 6200 & 27--35 & 2.3--3.3 & 5.6--7.3 & & & \\
    COPSS II & 17 & 2400 & 27--35 & 2.3--3.3 & 5.6--7.3 & & & \\
    mmIME & 2 & $5 + 15$ & 84--115$^a$ & & 1.0--1.7 & 2.0--3.1 & 3.0--4.5 & 4.0--5.9 \\
\enddata

\tablenotetext{a}{Not the whole frequency range was observed in both fields.}
\tablenotetext{b}{Primary targets were at a range of redshifts and the observed frequencies vary accordingly. The observed bandwidth for each target was 3.6 GHz.}
\end{deluxetable*}

Table~\ref{tab:surveys} summarizes the parameters of existing surveys used to measure the CO luminosity function or its moments. We list the number of fields observed, the survey area, the frequency coverage, and the corresponding redshift ranges in which CO(1-0) through CO(5-4) can be observed. We provide further implementation details and references for these projects in this section.

Four dedicated direct measurement surveys have been conducted. \citet{decarli+14} used the Plateau de Bure Interferometer to survey the 3 mm atmospheric window in a single pointing with a primary beam width of 55 arcsec. The survey resulted in secure detections of three objects, and identification of a number of additional candidates. \citet{walter+14} report constraints on the CO luminosity function derived from these candidates.

The ALMA Spectroscopic Survey in the Hubble Ultra Deep Field (ASPECS) consisted of two ALMA surveys. The pilot survey \citep{walter+16} conducted scans of ALMA bands 3 (3mm) and 6 (1.2mm) over a  0.9 arcmin$^2$ region in the Hubble Ultra Deep Field (HUDF). ASPECS pilot identified $\sim21$ line candidates (some corresponding to the same galaxy observed in different transitions), which \citet{decarli+16} used to provide luminosity function constraints.

The ASPECS Large Program \citep[hereafter referred to simply as ASPECS;][]{gonzalez-lopez+19} used the same spectral setup as the pilot survey, but surveyed a larger area covering 4.6 arcmin$^2$. These observations resulted in high-confidence identification of 12 objects in CO(2-1), a further five in CO(3-2), and one in CO(4-3) \citep{aravena+19}. \citet{decarli+19} use these objects along with a number of additional, lower confidence candidates to provide updated luminosity function constraints and \citet{uzgil+19} explored using intensity mapping techniques to constrain the luminosity function below the direct detection limit in the data set. 

The CO Luminosity Density at High-z \citep[COLDz;][]{pavesi+18} survey used the Jansky Very Large Array (JVLA) to search for CO emission in the Ka band (1cm), over two separate areas. The first was a 50.9 arcmin$^2$ region within GOODS-N, which was surveyed with shallower and non-uniform depth. The second was a 8.9 arcmin$^2$ region within COSMOS, which was $\sim 3$ times more sensitive but covered a smaller amount of area than the GOODS-N wide field. COLDz securely detected four objects in CO(1-0) and three in CO(2-1). Luminosity function constraints based on these objects and a large number of less secure candidates are reported by \citet{riechers+18}.

In addition, \citep{lenkic+20} used serendipitously detected secondary sources from the Plateau de Bure High-z Blue Sequence Survey 2 (PHIBSS2) to constrain the CO luminosity function. PHIBSS2 was primarily designed as a targeted CO line survey of redshifts 0.5--3.0 \citet{freundlich+19}. However, the large total volume covered by the survey allowed the identification of numerous secondary sources with no pre-selection, which can be used to constrain the luminosity function. PHIBSS2 consisted of 110 individual pointings, with a combined area of $\sim130$ arcmin$^2$. Because the survey's primarily purpose was targeted observations the frequency range for each pointing is much narrower (3.6 GHz), and the frequency window and corresponding redshift range vary from pointing to pointing. Integration times also vary by a factor of as much as $\sim$50. \citet{lenkic+20} identify 67 CO line candidates in this data set and estimate that $\sim 75$\% are likely to correspond to real objects.

Results from a handful of intensity mapping surveys are also available. The CO Power Spectrum Survey \citep[COPSS;][]{keating+15,keating+16} produced the first intensity mapping constraint on the CO luminosity function. COPSS used the Sunyaev-Zel'dovich Array (a subset of the Combined Array for Research in Millimeter-wave Astronomy) to constrain the CO(1-0) power spectrum at a wavelength of 1 cm. The first phase of the project (COPSS I) used archival data to set an upper limit on the power spectrum while the second phase (COPSS II) consisted of an optimized intensity mapping survey, collecting data over 17 independent pointings (each separated by $\gtrsim 1$ degree). Integration times varied from pointing to pointing with most fields receiving more than 100 hours of observation. \citet{keating+16} report a detection of the CO(1-0) power spectrum  at $z\sim2.6$ at 95\% confidence.

The Millimeter-wave Intensity Mapping Experiment (mmIME) consists of a series of surveys seeking to measure the shot component of the CO power spectrum. The first phase of the project targeted the 3mm spectral window using a combination of the archival ASPECS data and new Atacama Compact Array (ACA) observations over an additional 15 arcmin$^{2}$ \citep{keating+20arxiv}, resulting in a detection the CO shot power from a combination of CO(2-1), CO(3-2), and CO(4-3) at a 99.99\% confidence. The second phase, a Submillimeter Array survey of the 1mm spectral window, is ongoing.

\section{Model}\label{sec:model}
To simulate results of the measurements outlined in Section~\ref{sec:observables}, we use dark matter subhalo catalogs of the Illustris-TNG project's TNG300-1 simulation and a prescription for assigning CO luminosity to halos in order to create a set of 1000 light cones. We simulate the CO(1-0) emission line in the redshift range $0$ to $10$. For our fiducial survey, we extract objects in the redshift range $2.01<z<3.11$. This range corresponds roughly to the redshift range of CO(1-0) for the COLDz and COPSS observations and of CO(3-2) in \citet{keating+20arxiv}. It was chosen to exactly match the CO(3-2) redshift coverage of the ASPECS 3mm observations. Most current work in this field treats brightness ratios between CO(1-0) and higher-J lines as constants, therefore our CO(1-0) results can be directly compared to studies using other lines by re-scaling the luminosity axis. A line ratio of $L^\prime_{\rm CO(3-2)}/L^\prime_{\rm CO(1-0)}=0.42$ \citep{daddi+15} is commonly assumed \citep{decarli+19,lenkic+20}.

The Illustris-TNG simulations were a series of hydrodynamical simulations designed to study galaxy evolution in large cosmological volumes \citep{TNG1,TNG2,TNG3,TNG4,TNG5}. TNG300-1 is the largest volume simulation, with a comoving side length of 302.6 Mpc. The simulations includes 2500$^3$ dark matter particles of $5.9\times10^7$ M$_\odot$ and an additional 2500$^3$ ``gas'' particles of $1.1\times10^7$ M$_\odot$ which store baryonic information. The simulation's halo catalogs include $\sim14.5$ million objects at z=0, and record position, velocity, and halo mass, along with many baryonic/galaxy properties for each. Catalogs are provided for 100 snapshots covering a redshift range from 20 to 0. To construct light cones, we pick a line of sight direction and a random starting location in the simulation cube. We move along the line of sight, stepping through snapshots as the corresponding redshift increases, and extracting all halos in a square 500 arcmin$^2$ field up to redshift 10. We use the periodic boundaries of the box to include continuous large scale structure in our light cones by wrapping through the cube, and select lines of sight angled with respect to the cube faces so as to minimize repetition of the same structures. 

For our CO luminosity prescription, we follow \citet{li+16}.  Their model assigns star formation rates (SFRs) to each halo using the halo mass-SFR relationship from \citet{behroozi+13} and converts this to an IR luminosity. It then assigns CO luminosities using the empirical correlation between IR luminosity and CO(1-0) luminosity \citep{carilli+13,kennicutt+12,kennicutt98},
\begin{equation}\label{eq:IRCO}
    \log L_{\rm IR} = a \log L_{\rm CO} + b.
\end{equation}
Luminosities in this equation are in $L_\odot$ units and resulting CO luminosities can be converted to $L^\prime$ using equation~\ref{eq:lp}. A log-normal scatter of $\sigma$ is then applied to each halo luminosity to approximate astrophysical variations not accounted for in the model. For our fiducial model, we use $a=1.37$ and $b=-1.8$, and apply a scatter of $\sigma=0.35$ dex. We study the effects of altering our fiducial prescription in Section~\ref{sec:modeldependence}, and find that our main results are independent of choice of parameters.

Once halos have been selected and CO luminosities assigned, we generate a catalog containing the luminosity, sky position, and redshift of each object. We repeat this process 1000 times, generating new realizations of the CO luminosities each time, resulting in a set of 1000 catalogs. In order to generate smaller fields, we crop these large light cones to the required sizes.

At times throughout this paper, we need light cones free from cosmic variance in order to disentangle cosmic and Poisson variance effects. To generate an additional set of light cones with no large scale structure we create a catalog of all objects in the redshift range $z=2.01$ to $3.11$ from our original light cones. We then calculate the mean number of objects per light cone $\langle N\rangle$ and determine a number of objects $N$ to include in each new light cone by drawing from a Poisson distribution with mean $\langle N\rangle$. Finally we select $N$ galaxies from the combined catalog of all objects and include them in our new light cone, preserving their original redshifts and CO luminosities. This results in a set of light cones with the same redshift distribution and average luminosity function as our original simulations, but with no cosmic variance contribution to the distribution of these properties. We refer to these as Poisson light cones. 

\subsection{Independence of Large Scale Structure}

The finite size of the simulation limits the number of independent light cones that can be constructed. The TNG300-1 simulation has a volume of $8.6\times 10^6$ h$^{-3}$ Mpc$^{3}$. For redshifts $z=1.4$, $2.6$, $3.8$, and $5.0$, a box face corresponds to solid angles of $\Omega=16$, $8$, $6$, and $5$ square degrees and the length of the box corresponds to a redshift interval of $\Delta z=0.15$, $0.27$, $0.40$, and $0.56$. In this paper we use light cones ranging in solid angle from 1 to 500 arcmin$^2$. Table~\ref{tab:nlcs} lists the maximum number of light cones which can fit into the full simulation box for a number of solid angles. Because our light cones are constructed with a wide range of orientations to the original box, two light cones which both sample a particular point will still differ substantially overall and therefore the values in the table do not really represent a limit on the number of light cones.  

\begin{deluxetable}{c|ccccc}
\tablecaption{Number of independent light cones that can be drawn from the TNG300 simulation box\label{tab:nlcs}}
\tablehead{
    \colhead{} & \multicolumn{5}{c}{Number of Light Cones per TNG300 Volume} \\
    \colhead{Redshift} & \colhead{5am$^2$} & \colhead{15am$^2$} & \colhead{50am$^2$} & \colhead{150am$^2$} & \colhead{500am$^2$}}
\startdata
    1.0--1.7 & 2437 & 812 & 244 & 81 & 24 \\
    2.0--3.1 & 1379 & 460 & 138 & 46 & 14 \\
    3.0--4.5 & 1115 & 372 & 112 & 37 & 11 \\
    4.0--5.9 & 1010 & 337 & 101 & 34 & 10 \\
\enddata
\end{deluxetable}

\subsection{``True'' Properties of our Fiducial Model}

\begin{figure}
    \centering
    \includegraphics{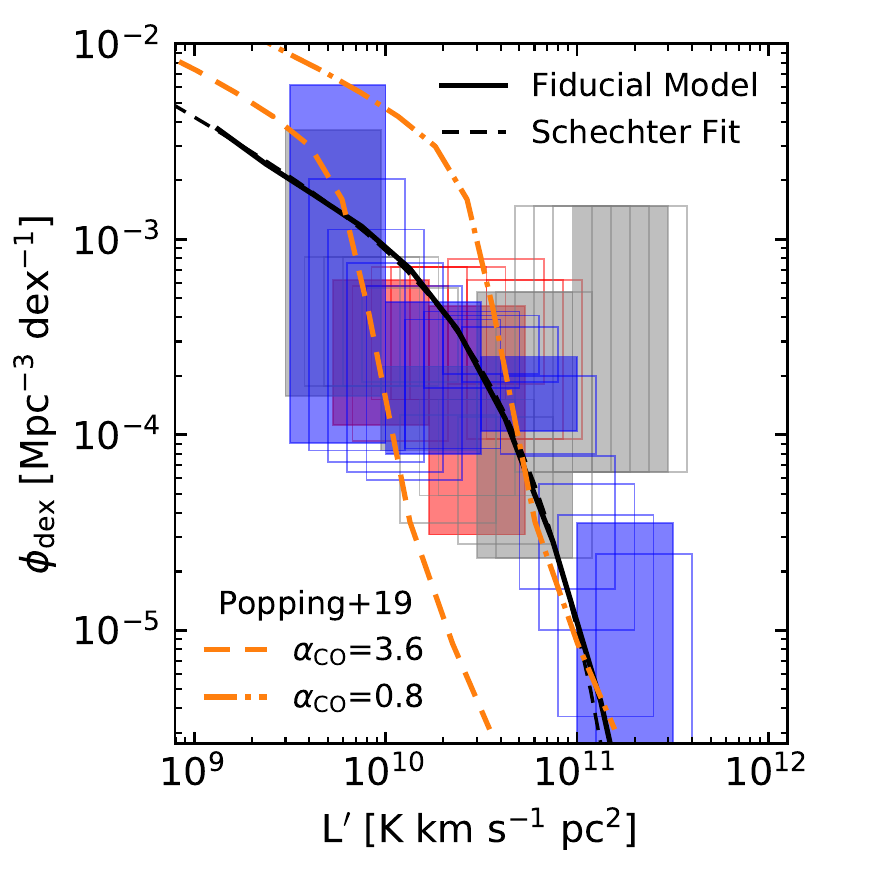}
    \caption{The luminosity function produced by our fiducial model over the redshift range $2.0<z<3.1$, measured in bins of 0.25 dex and the best fitting Schechter function are shown by the solid and dashed black lines respectively. The luminosity function constraints from the COLDz CO(1-0), ASPECS CO(3-2), and PHIBSS2 CO(3-2) direct line searches at $z\sim2$--$3$ are shown with blue, red, and gray colored boxes respectively. These studies have reported constraints in overlapping bins, producing covariant error boxes. We present independent luminosity ranges with filled boxes, showing only outlines for the remainder. CO(3-2) constraints are converted to CO(1-0) using a line ratio of 0.42 \citep{daddi+15}. The molecular gas mass function from the semianalytic model of \citet{popping+19}, converted to a luminosity function using $\alpha_{\rm CO}=3.6$ M$_\odot$ (K km s$^{-1}$ pc$^2$)$^{-1}$ and $0.8$ M$_\odot$ (K km s$^{-1}$ pc$^2$)$^{-1}$, is shown by the dashed and dash-dotted orange lines.}
    \label{fig:lfmodel}
\end{figure}

\begin{deluxetable*}{c|c|ccc}
\tablecaption{Schechter parameters of our model and a number of observational constraints\label{tab:fits}}
\tablehead{
    \colhead{Source} & \colhead{Redshift} & \colhead{$\log L_*$} & \colhead{$\log \phi_*$} & \colhead{$\alpha$}}
\startdata
    Model & 2.0--3.1 & 10.45 & -3.51 & -1.54 \\
    ASPECS CO(3-2)$^a$ & 2.0--3.1 & $10.98^{+0.20}_{-0.15}$ & $-3.83^{+0.13}_{-0.12}3$ & $-1.2$ (fixed) \\
    COLDz$^b$ & 2.0--2.8 & $10.7^{+0.63}_{-0.48}$ & $-3.87^{+0.67}_{-0.79}$ & $-0.92^{+0.91}_{-0.86}$ \\ \hline
    xCOLD GASS & 0.01--0.05 & $9.85^{+.10}_{-0.14}$ & $-2.89^{+0.19}_{-0.34}$ & $-1.13\pm0.05$ \\
\enddata
\tablenotetext{a}{Converted to CO(1-0) luminosity function using $L^\prime_{\rm CO(3-2)}=0.42L^\prime_{\rm CO(3-2)}$}
\tablenotetext{b}{COLDz reports 5th to 95th percentile confidence intervals. ASPECS and xCOLD GASS errors are 16th to 84th percentile}

\end{deluxetable*}

Our CO emitters are not drawn from a luminosity function, but rather the halo mass function output by the IllustrisTNG simulations convolved with our CO luminosity prescription. Therefore to define ``true'' luminosity functions and moments of our model against which we will compare our simulated observations, we combine our 1000 light cones into a single large catalog and measure the luminosity functions, moments, and other relevant physical quantities of the entire sample. 

The IllustrisTNG-300 simulation is large enough to capture scales up to the baryon acoustic oscillation scale \citep{TNG5}. We thus expect that our ensemble of light cones fairly samples the range of densities and should be minimally subject to cosmic variance, although limited sampling of the largest scales may result in some underestimation of these effects. Further, since the volume of our full set of light cones significantly exceeds the full simulation volume, we effectively produce many realizations of the same galaxies, sampling the full distribution of possible CO luminosities produced by scatter in our CO prescription.

Our ``true'' luminosity function at redshift $2.01$--$3.11$ is shown in black lines in Figure~\ref{fig:lfmodel}. Constraints from ASPECS \citep{decarli+19} at the same redshift range, PHIBSS2 at $z\sim2.2$ \citep{lenkic+20}, and COLDz at $z\sim2.4$ \citep{riechers+18} are also shown as boxes. These studies report constraints in a series of overlapping 0.5 dex bins, which are highly-correlated representations of the same data. We show independent measurements as filled boxes, and the remainder as open boxes. Our model is not designed to reproduce these measurements, but we note that there is reasonable agreement with most datasets over the range measured. Our model also produces comparable number counts to observations when sensitivity limits and completeness are accounted for. This means that where we find observational biases that are dependent on the underlying luminosity function, our model should be close enough to measurements that direct comparison of our simulated observations with those of real surveys is possible.

We perform a Schechter fit of this luminosity function following the procedure described in Section~\ref{sec:direct} and Appendix~\ref{appendix:fitting}, resulting in $L_*=2.81\times10^{10}$ K km s$^{-1}$ pc$^2$, $\phi_*=3.10\times10^{-4}$ Mpc$^{-3}$, and $\alpha=-1.54$. This Schechter fit is shown by the dashed line in Figure~\ref{fig:lfmodel} and compared to the values reported around $z\sim2.5$ by ASPECS and COLDz, and the $z\sim0$ fit from xCOLD GASS in Table~\ref{tab:fits}. Though our modeling does not require a luminosity function that is well-described by a Schechter function or any other parametric model, in practice we find good agreement between the fit and our simulated population over the luminosity range of our sample. For very rare objects (i.e. those found in fewer than $\sim 15$ percent of even our largest light cones) the true luminosity function falls off more slowly than the exponential part of the Schechter function. Our model results in $\alpha$ lower than the value found by \citet{saintonge+17} at $z\sim0$, the only redshift where this parameter is currently well constrained. However, we find in Section~\ref{sec:modeldependence} that a steep faint-end slope will only slightly alter the magnitude of biases seen in our simulations relative to real observations. The first moment of the luminosity function (mean brightness temperature) is 0.63 $\mu$K, corresponding to a mean molecular gas density of $4.8\times10^7$ M$_\odot$ Mpc$^{-3}$ for $\alpha_{\rm CO}=3.6$ M$_\odot$ (K km s$^{-1}$ pc$^2$)$^{-1}$. The second moment (shot power) is 502.54 $\mu$K$^2$ Mpc$^3$.

\section{Implications for the Luminosity Function}\label{sec:direct}

\begin{figure}
    \centering
    \includegraphics{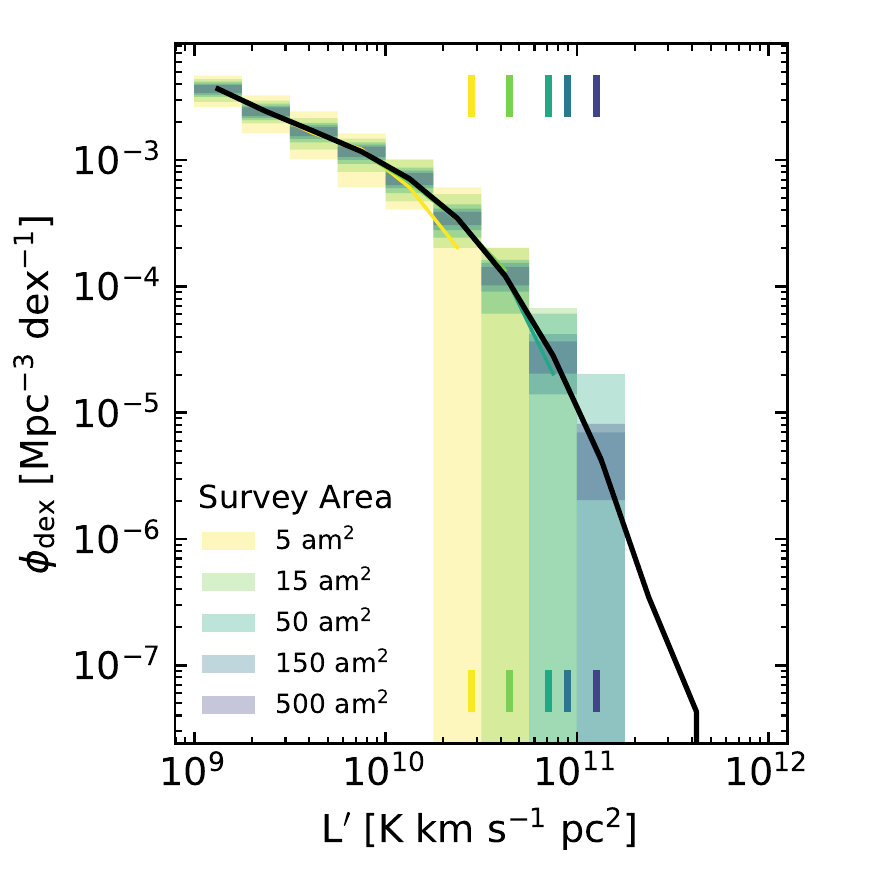}
    \caption{The true luminosity function (in Mpc$^{-3}$ dex$^{-1}$ units) for our fiducial model at median redshift 2.6 is shown as the black line. Colored lines and filled boxes represent the median and 16th-84th percentile spread in bins of 0.25 dex for a range of survey sizes. All survey sizes recover the true distribution up to a (field size dependent) maximum luminosity, where there is a sharp drop to zero as the expected number of brighter objects falls below $\sim0.7$ per survey volume. The thresholds where the median luminosity functions fall to zero are shown by thick vertical hashes at the top and bottom of the plot. These cutoffs happen at lower(higher) luminosities for smaller(larger) volumes, resulting in an apparent steepening of the luminosity function in small area surveys.}
    \label{fig:lf}
\end{figure}

\subsection{Estimation of the Luminosity Function}
We compute the luminosity function in each light cone from the object counts in luminosity bins of 0.25 dex width. In Figure~\ref{fig:lf} we show the median and 16-84th percentile range of the luminosity functions for light cones at a range of sizes from 5 to 500 am$^2$ (indicated by different color boxes). In general, the median value recovers the true luminosity function (black line) up to a cutoff luminosity, above which the survey becomes too small to reliably include brighter galaxies with low number densities. We indicate the location of this cutoff by vertical hashes at the top and bottom of the plot. For the smallest survey areas shown, the the cutoff coincides with the knee of the luminosity function.

This suggests that for small area surveys, the bright end of the luminosity function might appear to drop artificially rapidly. It could be argued that the luminosity functions of individual fields might still have enough filled bright bins to recover the shape of the bright end. However, we find that the same effect is present in the cumulative number density of galaxies, which is not subject to binning artifacts, so this is a real observational consequence of a small survey. Once the expected number of objects brighter than $L^\prime$ falls below a certain threshold for a given survey size, the majority of surveys will not detect any such objects\footnote{Assuming that the number of bright galaxies per field is approximately Poisson distributed, this threshold is $V_{\rm obs}\langle n(>L^\prime) \rangle \sim 0.7$}. This implies that accurately identifying the knee of the luminosity function requires a survey volume for which at least a few objects brighter than the turnover luminosity are expected. Otherwise, artificial cutoffs introduced by the survey itself may be misidentified as real changes in shape. 

\begin{figure*}
    \centering
    \includegraphics[width=\textwidth]{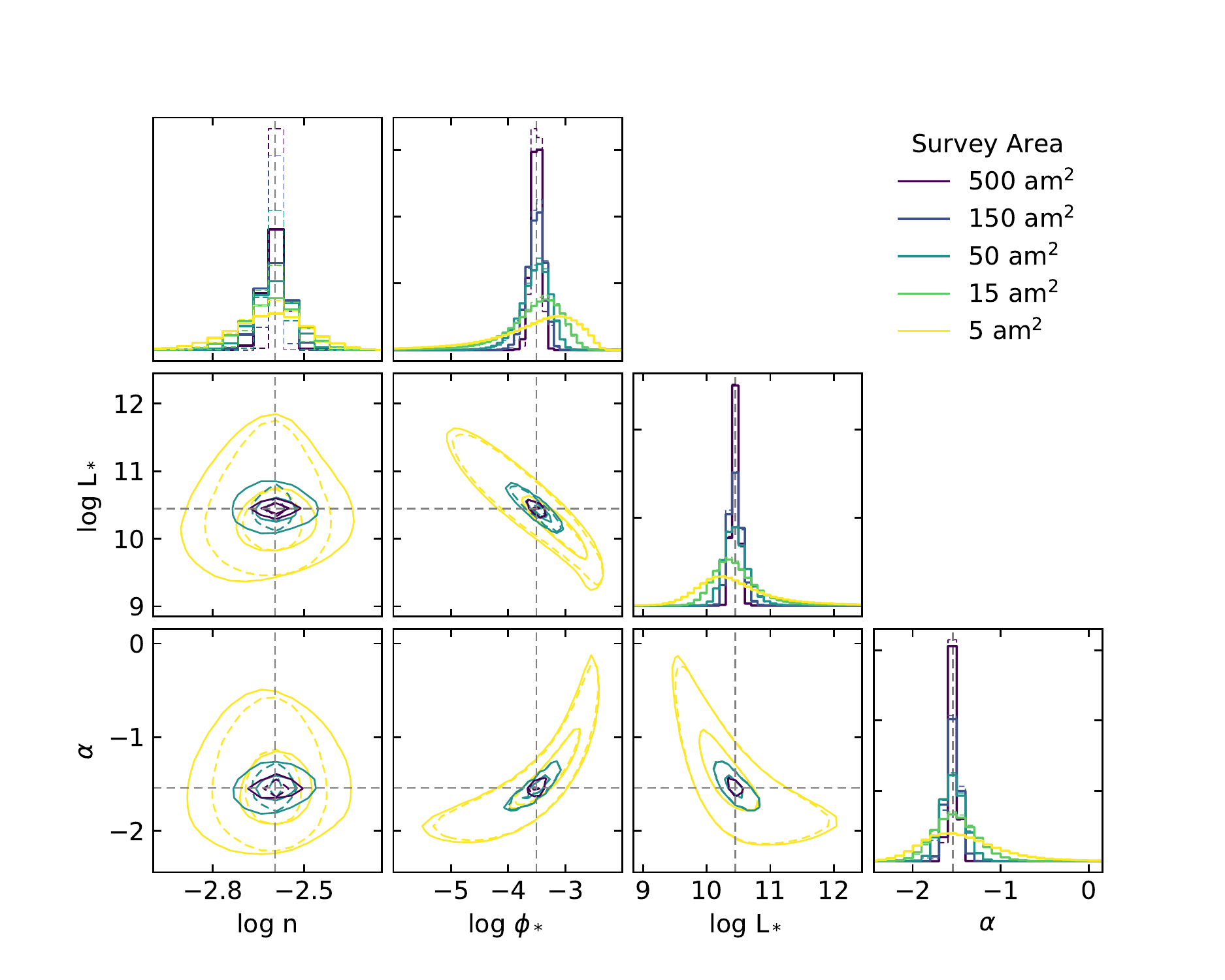}
    \caption{Combined distributions of fitted Schechter parameters for all 1000 light cones. We show the normalization parameterized using both $\phi_*$ and $n=\int \phi(L^\prime)dL^\prime$. In the upper plot of each column the one-dimensional probability densities for each parameter are shown for five different survey areas. Gray dashed lines show the true value of each parameter. Note that small surveys show a bias towards low $L_*$ and high $\phi_*$ in addition to wider distributions. The remaining plots show contours containing 39.9 and 86.5\% (corresponding to 1$\sigma$ and 2$\sigma$ confidence intervals in two dimensions) of the fitted values in pairs of parameters. Dashed histograms/contours are the results for light cones with no large scale structure, which show slightly tighter two dimensional constraints, but virtually identical one dimensional results for $\phi_*$ and the shape parameters $L_*$ and $\alpha$. For $n$, which removes the shape dependence from the normalization term, a clear increase in variance can be seen when cosmic variance is included.}
    \label{fig:mcmccorner}
\end{figure*}

To assess how this affects parametric fits of the luminosity function, we used a Markov Chain Monte Carlo (MCMC) fitting procedure to fit Schechter functions to each of our light cones. The fitting is implemented using the emcee package \citep{foreman-mackey+13}, and is described in detail in Appendix~\ref{appendix:fitting}. Figure~\ref{fig:mcmccorner} shows the combined parameter distributions for all 1000 light cones, for a range of light cone sizes. For fields in the 5-15 arcmin$^2$ range (shown in yellow and light green), the one dimensional distribution for $L_*$ becomes biased toward small values, as expected from the cutoff effect noted above.

The magnitude of this effect is dependent upon the underlying luminosity function. In particular, luminosity functions where the normalization at $L_*$ is higher (and thus more objects above $L_*$ are expected) will produce less severe biases, while luminosity functions with lower normalization will exacerbate the issue. This means that the size of field needed to constrain the turnover of the luminosity function is dependent on the luminosity function itself.

\begin{figure}
    \centering
    \includegraphics{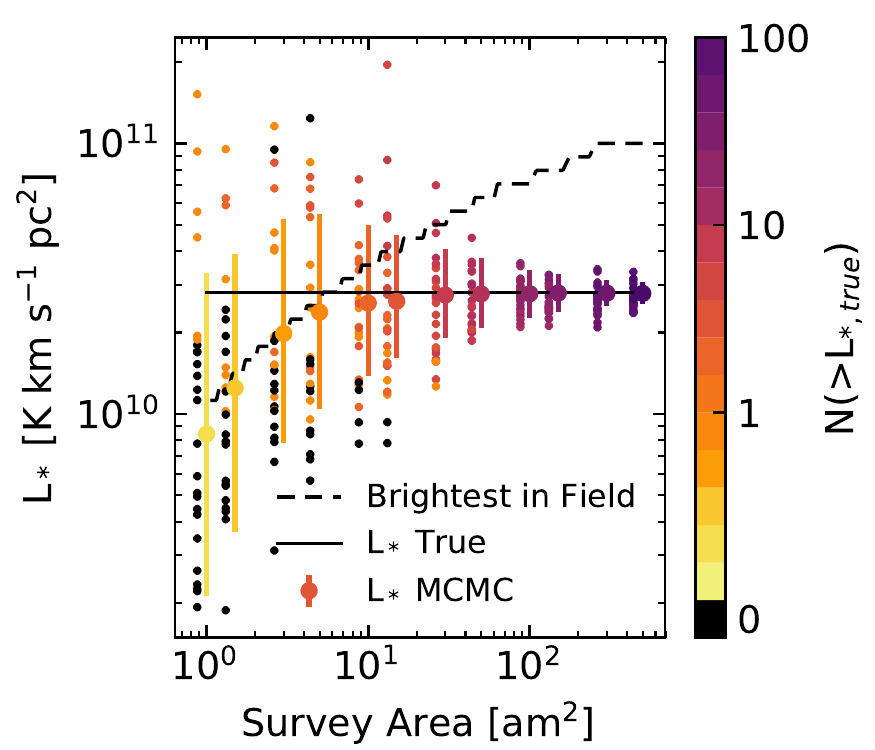}
    \caption{Fitted values of $L_*$ from our MCMC parameter fitting are shown as a function of survey area. The true value of $L_*$ is marked with a horizontal black line. Large points with vertical error bars show the median and 16th to 84th percentile spread in the fitted value of $L_*$ in the ensemble of light cones, and are color coded by the expected number of galaxies with luminosity greater than $L_*$.
    Small points (offset to the left) show the fits for a sample of 25 individual light cones, and are colored according to the number of galaxies with luminosity greater than $L_*$ found in the individual survey volume. The dashed black line shows the brightest object likely to appear in the survey volume, defined as the luminosity above which the cumulative number equals one.}
    \label{fig:ngtls}
\end{figure}

Because the shape of the luminosity function is not known in advance, caution must be exercised when using surveys of small volumes to constrain the luminosity function. Figure~\ref{fig:ngtls} shows the distribution of fitted values of $L_*$ as a function of survey area (large points with error bars). The dashed black line shows the brightest galaxy likely to appear in a field of a given area, and where this value drops below the true value of $L_*$ (solid black line), the fitted value of $L_*$ tends to be biased below the true value. For our simulated light cones this happens for fields smaller than $\sim10$ arcmin$^2$. Considering individual light cone realizations (small points), those that have zero galaxies with $L>L_*$ have fitted values of $L_*$ below the true value and those with one or more such galaxies typically infer high values.

A similar effect has been encountered in the optical/IR surveys used to constrain the rest frame 1500 \si{\angstrom} luminosity function at high redshift. Early constraints at $z\gtrapprox4$ found evidence for for a decline in characteristic luminosity luminosity $L_{*,UV}$ \citep[e.g.][]{bouwens+08,su+11}. Later analysis using larger area datasets covering hundreds of arcmin$^2$ did not replicate this evolution, finding instead that a steepening of the faint-end slope better fit the data  \citep{finkelstein+15}. \citet{bouwens+15} find that the apparent evolution of $L_{*,UV}$ was partly due to small areas used in previous analyses poorly constraining the bright end of the luminosity function, especially in the presence of steep faint-end slopes. This has important consequences for models of galaxy evolution, as a changing characteristic luminosity lends to different physical interpretations than changes in the faint-end slope \citep[e.g.][]{jaacks+12,somerville+12}.

\subsection{Cosmic Variance in Luminosity Function Constraints}\label{sec:cvdirect}

The fields used for CO luminosity function studies have all contained 15 or fewer objects per redshift window, compared to hundreds or thousands used in typical studies of the UV luminosity function at comparable redshifts \citep[e.g.][]{arnouts+05,reddy+09}. Results for the CO luminosity function have therefore dismissed cosmic variance as negligible in comparison to Poisson variance, owing to the small sample sizes involved.

\begin{figure*}
    \centering
    \includegraphics[width=\textwidth]{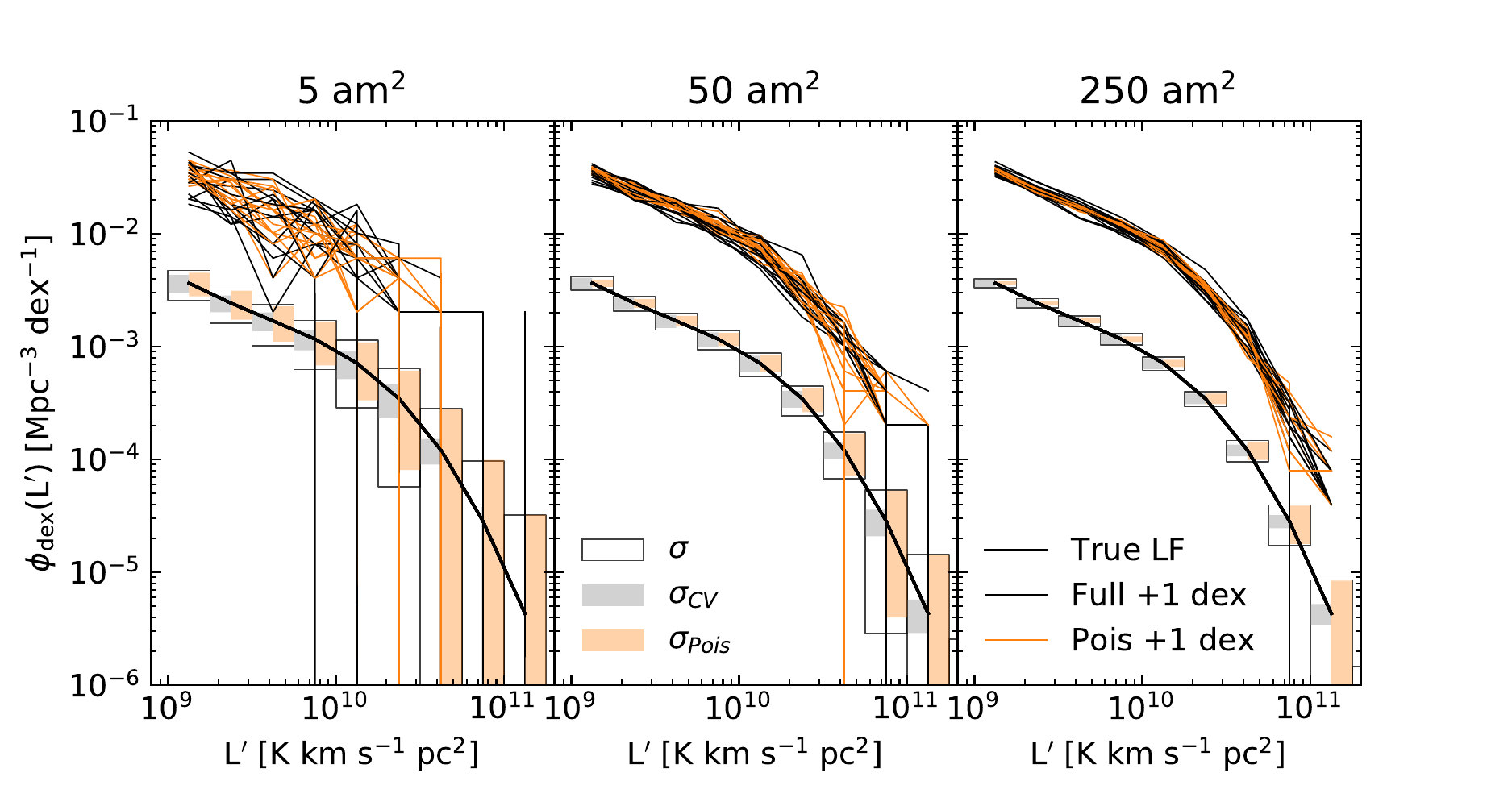}
    \caption{The true luminosity function (heavy black line) is shown with black error boxes showing the standard deviation in $0.25$ dex bins for light cones of 5, 50, and 250 am$^2$ survey area. The colored areas in each error box show the breakdown of the standard deviation into Poisson variance (orange) and cosmic variance (gray) contributions. Also shown, offset upwards by 1 dex, are the measured luminosity functions for 20 light cones drawn from simulations with (black) and without (orange) large scale structure.}
    \label{fig:lfindiv}
\end{figure*}

To validate this, we use our Poisson (large scale structure-free) light cones. Figure~\ref{fig:lfindiv} shows the measured luminosity functions for 20 light cones with (black) and without (cyan) cosmic variance, in 5, 50, and 250 arcmin$^2$ areas. In the smallest areas, the Poisson light cone luminosity functions show similar behavior to the full simulation light cones, with counts in each bin largely uncorrelated. In volumes this small, the counts in all bins are sufficiently low that Poisson variance dominates.

For the larger areas the Poisson light cones show reduced spread. However, while the bins of the Poisson luminosity functions are uncorrelated, in the luminosity functions where large scale structure is taken into account there is a clear covariance between bins. This is a cosmic variance effect -- as different light cones probe underdense(overdense) regions, the number of objects at all luminosities falls(rises). In the 50 am$^2$ area, the counts for the Poisson light cones still show appreciable variation, indicating that Poisson variance continues to be important at all luminosities, and dominates at the bright end. However, as the area increases the magnitude of the Poisson variance falls faster than the cosmic variance contribution and more bins become cosmic variance dominated. 

\begin{figure*}
    \centering
    \includegraphics[width=\textwidth]{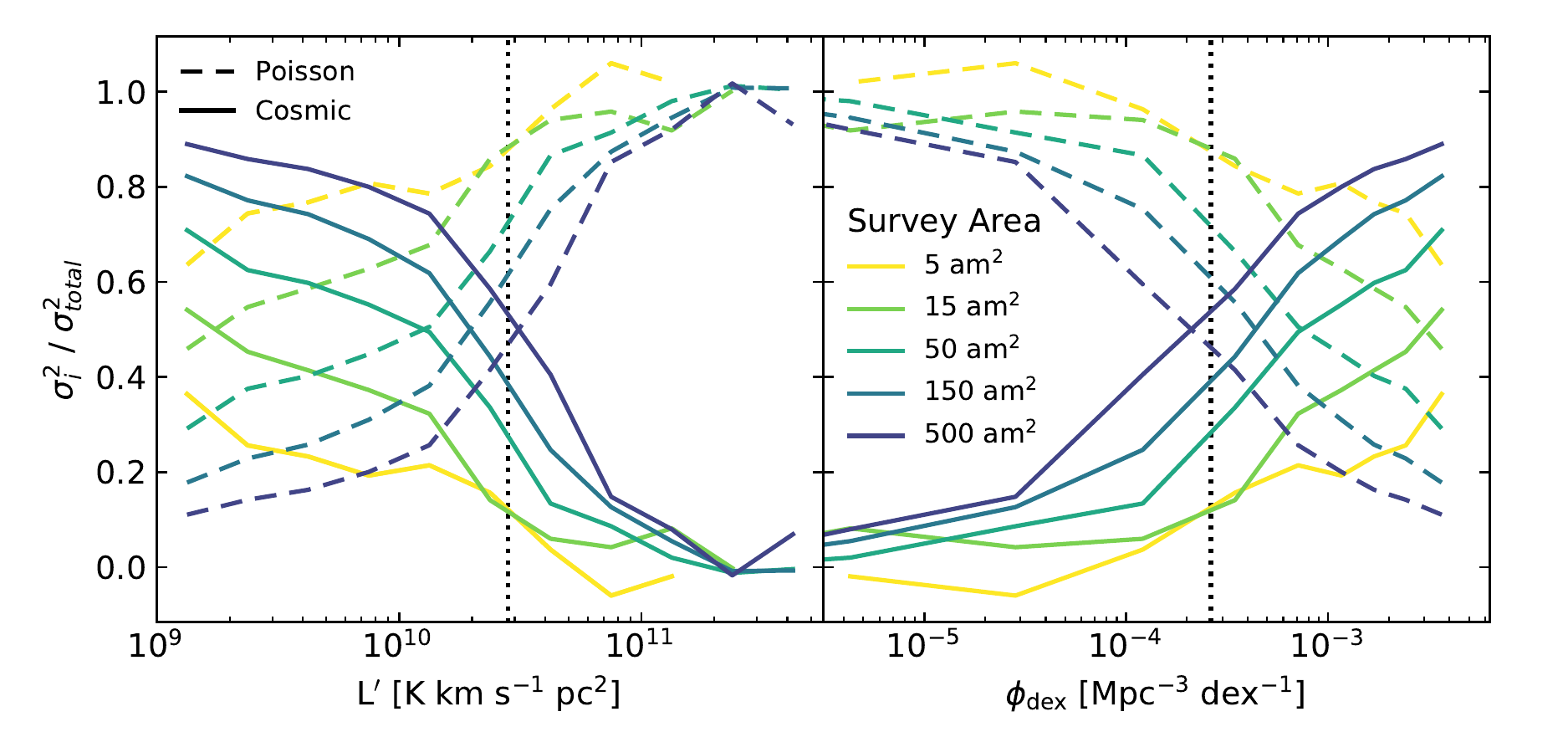}
    \caption{Breakdown of the sample variance into its cosmic (solid line) and Poisson (dashed line) components for a range of survey sizes shown in different colors. In the left panel this is shown as the fractional contribution of each type of variance in $0.25$ dex luminosity bins, and in the right panel it is shown as a function of the number density of objects. The dashed vertical lines indicate the characteristic luminosity $L_*$ (left) and normalization $\phi(L_*)$ (right) of our model, where $L_*=2.81\times10^{10}$ K km s$^{-1}$ pc$^2$}
    \label{fig:cv}
\end{figure*}

The left panel of Figure~\ref{fig:cv} shows the relative contribution of Poisson (dashed lines) and cosmic (solid lines) variance to the sample variance in each luminosity bin, for a range of survey sizes. 
We separate the cosmic variance from the Poisson variance in each luminosity function bin using equation~\ref{eq:cv}. The total sample variance is measured from the 1000 full light cones and the Poisson variance is computed based on the mean number of sources per bin. For the smallest areas, Poisson variance is the dominant contribution in all bins, and contributes over 90 percent for bins above $L_*$ ($2.81\times10^{10}$ K km s$^{-1}$ pc$^2$, shown by the vertical line). As surveys become larger, cosmic variance grows in importance for low luminosity bins, reaching $\sim90$ percent in the largest surveys. However, even for largest areas, Poisson variance begins to outweigh cosmic variance around $L_*$.

The exact behavior of the curves on the left of Figure~\ref{fig:cv} depends on underlying luminosity function, because it is number density, not luminosity, that sets the degree of variance. In the right panel of Figure~\ref{fig:cv} we map each luminosity bin to its corresponding value of $\phi$ using the true luminosity function for our model. Although the degree of cosmic variance may have some secondary dependence on the exact input model, this representation should allow reasonable, model independent comparisons. Note that when the line of sight length, $d$, of the survey is greater than the characteristic scale of clustering, both the cosmic and Poisson variance should scale as $\sim 1/d$ \citep{driver+10}, thus when comparing different surveys these results should be fairly insensitive to the exact length of the redshift interval.

We find that below $2\times10^{-4}$ Mpc$^{-3}$ dex$^{-1}$ Poisson variance is the main contributor to the variance for all survey sizes considered. Above this density, cosmic variance begins to dominate for large light cones. The ASPECS CO(3-2) luminosity function measurements cluster around $10^{-3.6}$ Mpc$^{-3}$ dex$^{-1}$. Figure~\ref{fig:cv} shows that for a 5 arcmin$^2$ field size cosmic variance is less than 20\% of the total sample variance. This confirms that cosmic variance is of secondary importance for ASPECS' constraints on \textit{individual bins} of the luminosity function. However, for ASPECS lower redshift CO(2-1) results, some bins are as high as $10^{-2.5}$ Mpc$^{-3}$ dex$^{-1}$. For such bins cosmic variance approaches 40\% of the total sample variance and is large enough that it should not be neglected. For surveys comparable in size to the COLDz wide field, Poisson variance is the primary contribution below $\sim7\times10^{-4}$ Mpc$^{-3}$ dex$^{-1}$.  Many of the COLDz data points fall in the range $10^{-4}$ to $10^{-3}$ Mpc$^{-3}$ dex$^{-1}$ where cosmic variance contributes 15--60\% in a 50 arcmin$^2$ field. We note that COLDz and ASPECS present their luminosity functions in bins of 0.5 dex rather than 0.25 dex used here. We explored the effects of adopting larger bins and find that our results are unchanged.

In Figure~\ref{fig:mcmccorner} we show the results of Schechter fits to our Poisson light cones with dashed lines. The $2\sigma$ contours for $L_*$, $\phi_*$, and $\alpha$ become slightly tighter after removing the effects of cosmic variance, but otherwise we see no discernible effect on our fits for these parameters. At the faint end, the effect of cosmic variance is to jointly shift bins up or down, without affecting the overall slope, as can be seen in the individual light cone luminosity functions of Figure~\ref{fig:lfindiv}. Thus cosmic variance effects should leave $\alpha$ unaltered. Furthermore, Poisson variance is the primary source of uncertainty for bins at or above $L_*$ even for the largest survey considered here, so cosmic variance should have a limited effect on the fitted value of $L_*$. 

The parameter where we would expect to see the effect of cosmic variance most strongly is the normalization $\phi_*$, but in the formulation given in equation~\ref{eq:schechter}, $\phi_*$ contains information about both the shape and normalization of the luminosity function, and the dependence on the shape parameters hides the effect of cosmic variance. 
Figure~\ref{fig:mcmccorner} also shows the joint distribution of the mean number density of all CO-emitting objects, defined as
\begin{equation}\label{eq:schn}
     n = \phi_* \int (L^\prime/L_*)^\alpha \exp (-L^\prime/L_*) dL^\prime \,.
\end{equation}
This reparameterization separates the shape and normalization, and we do find a significant widening of the contours in the $n$ direction when cosmic variance is included, even for the smallest survey areas. This highlights the important fact that integral quantities relating to the luminosity function are more subject to cosmic variance than individual bins. We explore this further in the following section.

\section{Implications for Moments of the Luminosity Function}\label{sec:moments}

The cosmic molecular gas density is an integrated measurement, corresponding to the first moment of the luminosity function times $\alpha_{\rm CO}$. Direct detection efforts seek to constrain this quantity by summing over the detected objects. Intensity mapping naturally measures moments of the luminosity function, with clustering power proportional to the first moment and the shot power proportional to the second. In this section we explore measurements of the first and second moment of the luminosity function.

\begin{figure}
    \centering
    \includegraphics{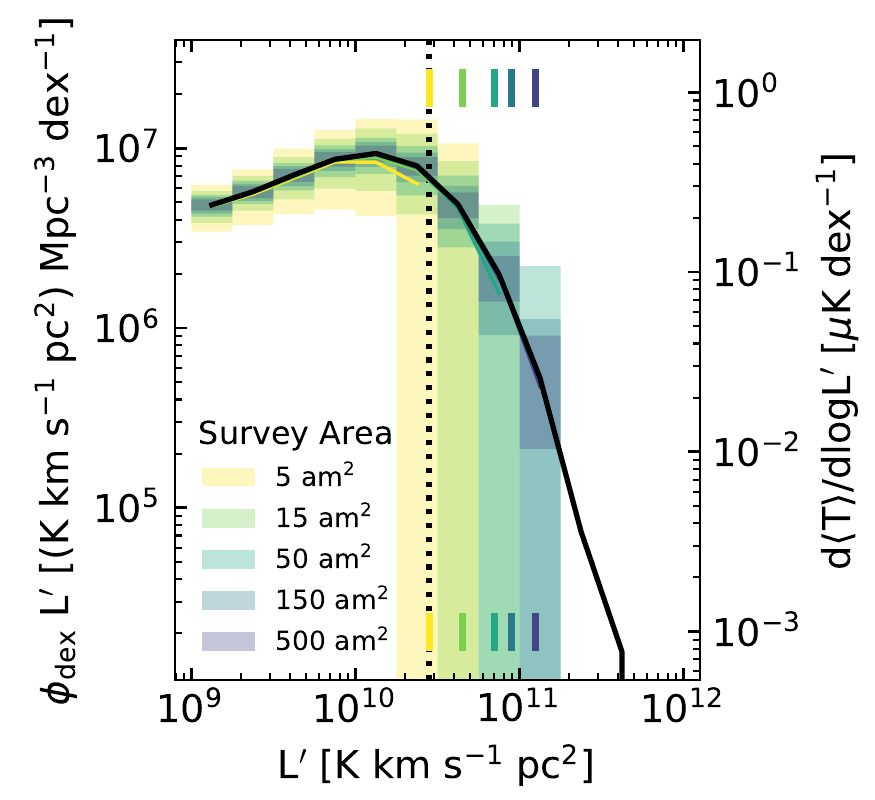}
    \includegraphics{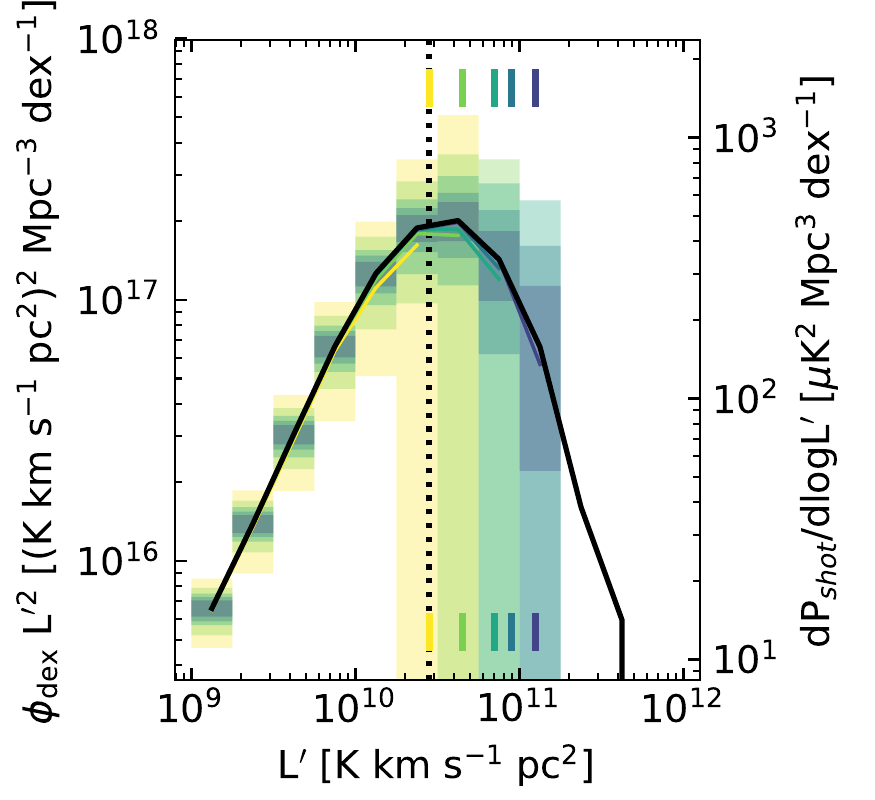}
    \caption{Differential contributions from luminosity bins of 0.25 dex to the first (top) and second (bottom) moments of the luminosity function. The left vertical axis give units in terms of the raw moments of the luminosity function while the right axis is converted to differentials of the corresponding observed quantities. Shown in black is the true contribution from each bin over the full sky. The thin lines and colored boxes show the median and 16th-84th percentile range for fields of a variety of sizes. The dashed vertical lines indicate the characteristic luminosity of our model, $L_*$. At a given field size, the median shows a cutoff, indicated by the vertical hashes at the top and bottom of each panel, above which contributions are not recovered, because brighter objects have too low a number density. This cutoff biases the measured moments of the luminosity low, especially for the smallest surveys, where even bins near the peak appear to have zero contribution.}
    \label{fig:lfmom}
\end{figure}

The luminosity weighting of the moments gives added importance to the bright end, where low number densities create the largest uncertainties. Figure~\ref{fig:lfmom} shows the median and 16th-84th percentile range of the quantities $\phi(L^\prime)\times L^\prime$ and $\phi(L^\prime) \times L^{\prime 2}$ as a function of $L^\prime$ for a range of survey sizes. These are the quantities integrated to determine the first and second moments of the luminosity function, and the right axes give them in units of differential mean brightness temperature ($\mu$K dex$^{-1}$) and power ($\mu$K$^2$ Mpc$^3$ dex$^{-1}$). A wide range of luminosities contribute to the first moment, while the $L^{\prime 2}$ weighting of the second moment creates a sharp peak near $L_*$. 
As shown in figure~\ref{fig:lf}, there is a survey size-dependent cutoff luminosity above which most light cones have no bright objects. For measurements of the moments of the luminosity function, the effect of this cutoff is more severe. The cutoff luminosity encroaches on the bins contributing the most to the first moment and reaches below the dominant bins for the second moment in the smallest survey areas. Small surveys can therefore be expected to underestimate both moments of the luminosity function, with the most severe effect seen in the second moment.

\subsection{Estimation of the Moments}\label{sec:estmoments}

We calculate the moments of our light cones by summing over all objects down to $L^\prime=10^9$ K km s$^{-1}$ pc$^2$.
Figure~\ref{fig:lfmom} shows that this cutoff excludes very little of the total contribution to the second moment, but may underestimate the first moment. In reality, the CO luminosity function is expected to drop off at some point, since baryonic processes are expected to become less efficient in halos below some mass cutoff \citep{pullen+13}. However, the location of this cutoff is not known. Integrating the Schechter fit of our fiducial model down to $L^\prime=10^{7.5}$ K km s$^{-1}$ pc$^2$, we find that our cutoff misses approximately 23\% of the first moment, with even lower luminosities contributing negligibly. Similarly, running our model itself including galaxies down to $L^\prime=10^{7.5}$ K km s$^{-1}$ pc$^2$ suggests 29\% of the first moment is missed by the higher cutoff (the small discrepancy between the two approaches is due to a slight upturn in the model luminosity function below the range for which we perform our Schechter fitting). These corrections depend on the faint-end slope of the luminosity function, with shallower slopes resulting in smaller contributions below $10^9$ K km s$^{-1}$ pc$^2$. For direct detection surveys, this is irrelevant, as current instruments lack the sensitivity to probe these lower luminosities. For intensity mapping, the clustering power is proportional to the first moment, so including these galaxies may alter the uncertainties somewhat (shot power is unaffected as the second moment is insensitive to these galaxies). Section~\ref{sec:modeldependence} explores the effects of changing the faint-end slope of the luminosity function, which has much the same effect as including fainter galaxies, and finds that our results are unchanged. In any case, since objects above $10^9$ K km s$^{-1}$ pc$^2$ contribute $\sim$75\% of the first moment, we assume the effect of fainter objects is smaller than uncertainties due to model selection and do not consider them further. 

\begin{figure*}
    \centering
    \includegraphics[width=\textwidth]{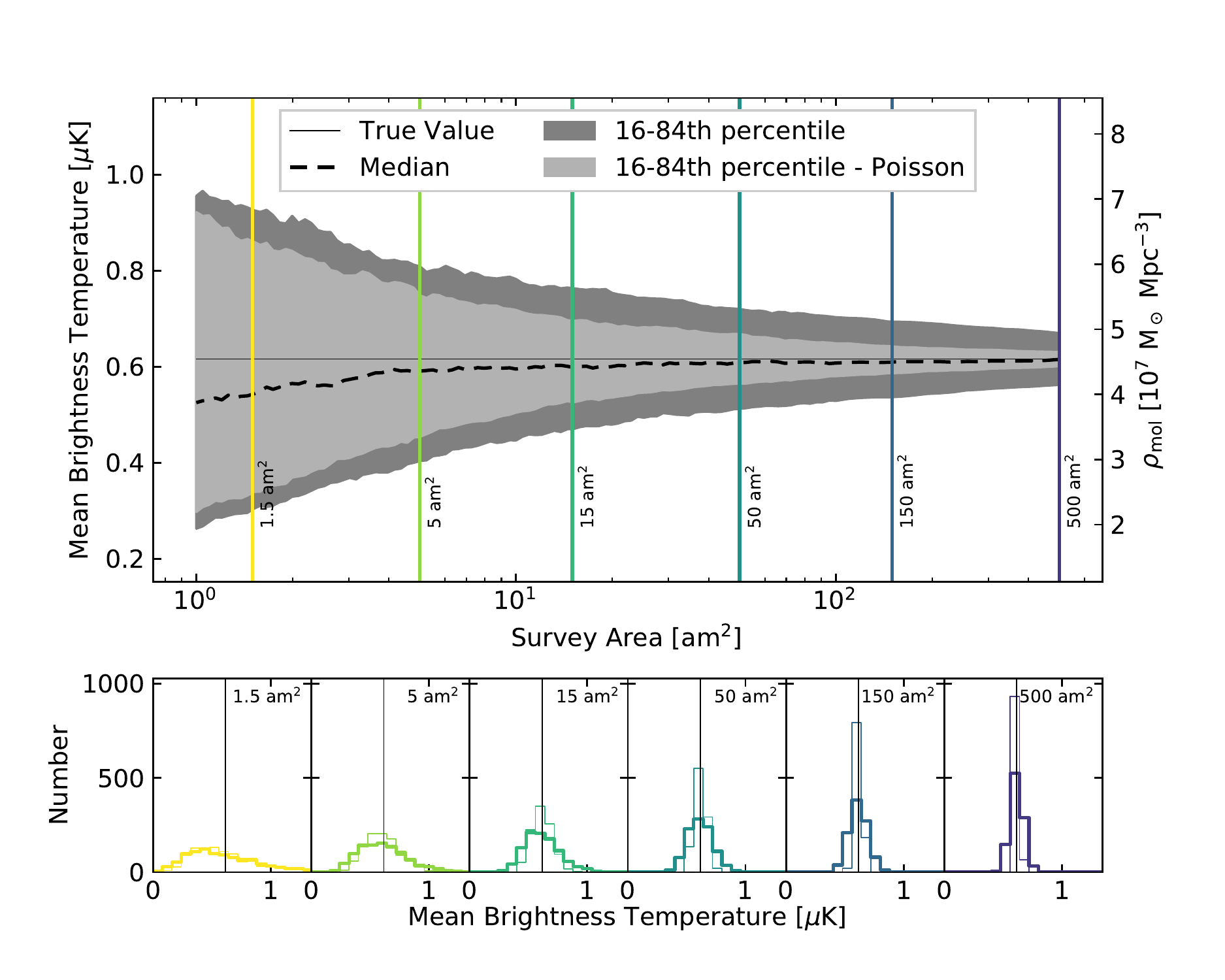}
    \caption{Top: Median (dashed line) and 16th-84th percentile range (filled area) of first moment measurements as a function of survey area for light cones with (dark gray) and without (light gray) large scale structure. The solid horizontal line indicates the true value implied by our model. Colored vertical lines correspond to areas for which we show the distribution of moments for all 1000 light cones in the bottom panels. On the right axis we show units of mean molecular gas density, computed using Equation~\ref{eq:Trho} and $\alpha_{\rm CO}=3.6$ M$_\odot$ (K km s$^{-1}$ pc$^2$)$^{-1}$.
    Bottom: Distribution of mean brightness temperatures for light cones with (thick lines) and without (thin lines) large scale structure.}
    \label{fig:TvA}
\end{figure*}
\begin{figure}
    \centering
    \includegraphics{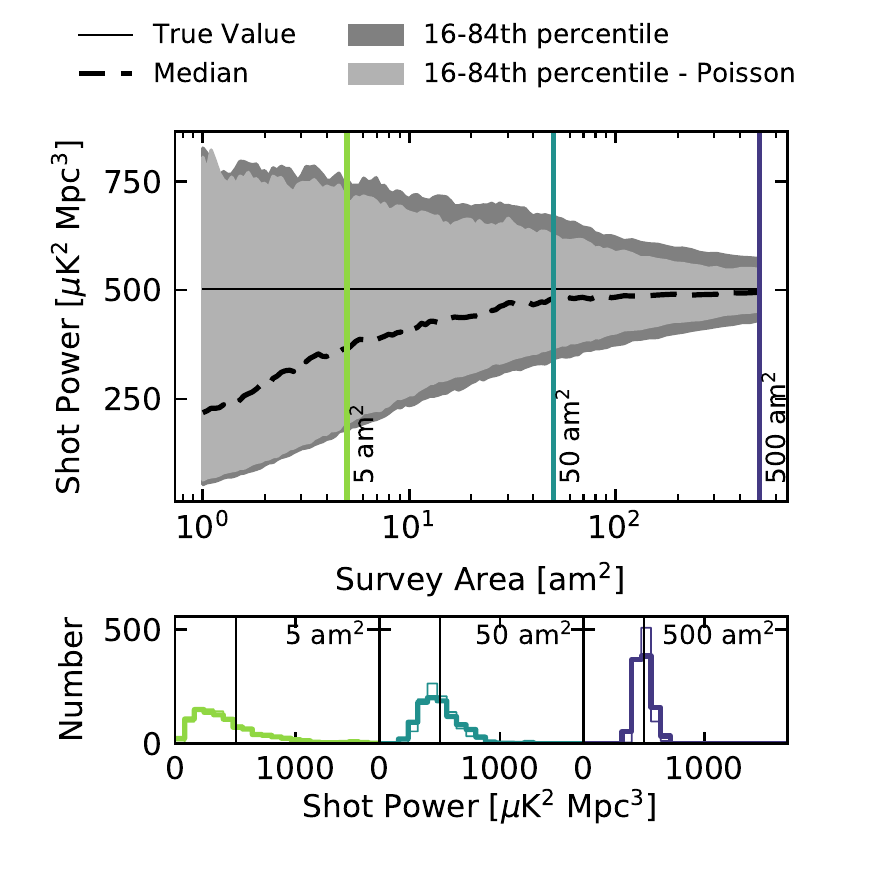}
    \caption{Top: Median (dashed line) and 16th-84th percentile range (filled area) of second moment measurements as a function of survey area for light cones with (dark gray) and without (light gray) large scale structure. The solid horizontal line indicates the true value implied by our model. Colored vertical lines correspond to areas for which we show the distribution of moments for all 1000 light cones in the bottom panels. Bottom: Distribution of mean brightness temperatures for light cones with (thick lines) and without (thin lines) large scale structure.}
    \label{fig:SvA}
\end{figure}

In the upper panels of Figures~\ref{fig:TvA} and~\ref{fig:SvA} we show the medians and distributions of the measured mean brightness temperature (corresponding to the first moment) and measured shot power (corresponding to the second moment) as a function of survey area. As expected, small area surveys tend to underestimate the true moments. For the mean brightness temperature, and by extension $\rho_{\rm mol}$, this effect is relatively small, just 8\% at 5 am$^2$. As noted previously, the left panel of Figure~\ref{fig:lfmom} helps explain why - galaxies well below $L_*$ that are numerous in fields of all sizes contribute significantly to the first moment while bright galaxies missed by small surveys contribute only a fraction of the total value.

\begin{figure}
    \centering
    \includegraphics{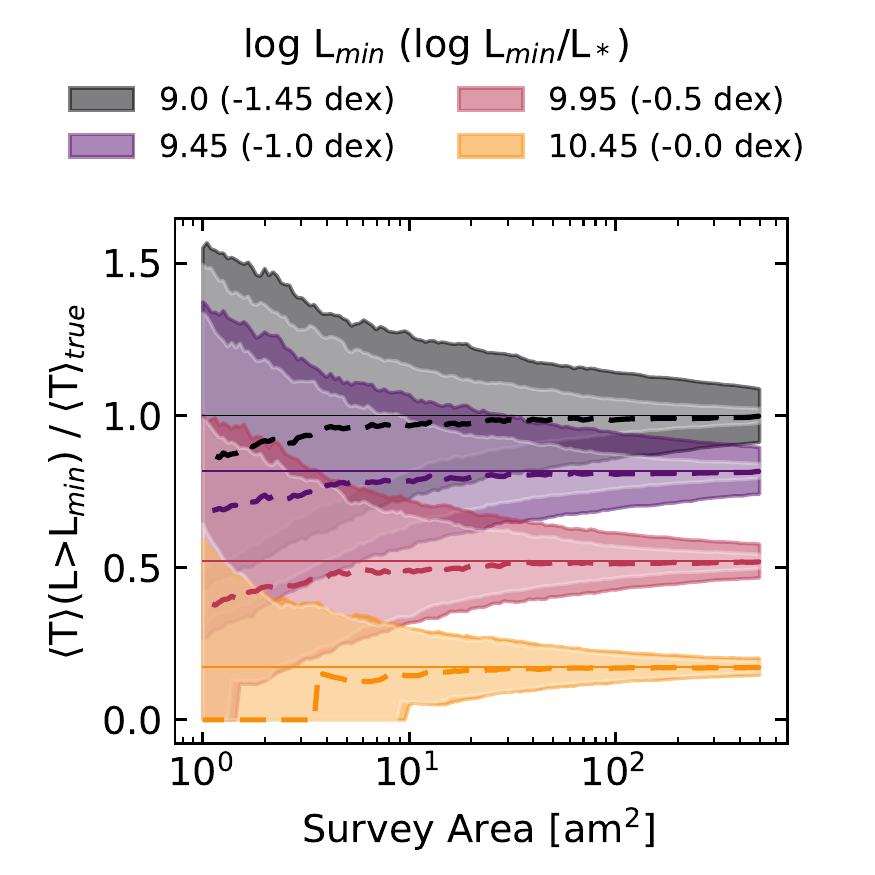}
    \caption{True value (solid lines), median value (dashed lines), and 16th-84th percentile range (filled areas) of of the measured mean brightness temperature of objects above a series of $L^\prime$ cutoffs ranging from $10^{9.00}$ K km s$^{-1}$ pc$^2$ (1.45 dex below $L*$ for this model) to $10^{10.45}$ K km s$^{-1}$ pc$^2$ ($L*$). The values are normalized by the true mean brightness temperature of all objects brighter than $10^{9.00}$ K km s$^{-1}$ pc$^2$ ($\langle T\rangle_{\rm true}=0.63$~$\mu$K). Light filled regions indicate the Poisson contribution to the range, while darker regions include cosmic variance.}
    \label{fig:cut}
\end{figure}

In practice direct detection surveys have been limited in depth to around the knee of the luminosity function, and these surveys measure only the contribution to the mean brightness temperature from objects above the detection threshold \citep{decarli+19}. To represent this effect, we recompute our moment statistics using catalogs truncated at a range of luminosities. Figure~\ref{fig:cut} shows the median and range of truncated first moments for a number of cutoffs, normalized by the true mean brightness temperature of our model. The primary effect of the cutoff is to reduce the recovered fraction of the mean brightness temperature. For small surveys and higher luminosity cutoffs there is also a small increase in the discrepancy between the median and true values (dashed and solid lines). 

An ASPECS-like (5 arcmin$^2$) survey that probes 1 dex below $L_*$ can be expected to measure between 61 and 136\% (16th-84th percentile range) of the mean brightness temperature contribution of objects bright enough to be detected, with a median value of 95\%. This corresponds to 50 to 111\% of the full mean brightness temperature, with a median of 77\%. On the other hand, for a cutoff at $L_*$, the median survey recovers only 13\% of the full mean brightness temperature, with a range from 0 to 34\%.

Figure~\ref{fig:SvA} shows that small surveys face a much more severe downward bias when constraining the shot power (second moment). The median light cone measures a second moment substantially lower than the true second moment at a wide range of survey sizes. This is because a handful of bright galaxies around the knee of the luminosity function contribute the majority of the shot power, and small surveys do not cover enough area to reliably recover their contributions. 

\subsection{Cosmic Variance in Moment Constraints}\label{sec:cvmoments}

Figures~\ref{fig:TvA} and~\ref{fig:SvA} show the distribution of moment measurements for light cones with and without large scale structure. 

For the first moment, cosmic variance makes a greater contribution to the overall spread of our measurements, even down to very small survey areas. Inserting the variance in mean brightness temperatures for our full and Poisson light cones into Equation~\ref{eq:cv} we find that the cosmic variance is equal to 45\% of the total sample variance for a 5 arcmin$^2$ survey. For larger surveys this rises to 65\%, 77\%, 87\%, and 92\% of the total for 15, 50, 150, and 500 arcmin$^2$ areas respectively. When cosmic variance is not accounted for this yields error bars that are too small by 25\% for the 5 arcmin$^2$ survey, and the larger surveys underestimate their errors by 41\%, 52\%, 64\%, and 72\% respectively. This suggests that the error bars in plots of the redshift evolution of $\rho_{\rm mol}$ have been underestimated. We explore how accounting for these errors affects this constraints on this evolution in the next section.

For the second moment, the inclusion of large scale structure adds only minimally to the variance. For our 500 arcmin$^2$ survey, cosmic variance accounts for 33\% of the sample variance, leading to error bars too small by 19\%. This follows from the results of Section~\ref{sec:cvdirect} where we found that for the galaxies around $L_*$, to which the second moment is most sensitive, Poisson variance outweighs cosmic variance for all survey sizes considered.

\section{Redshift Evolution}\label{sec:rs}

\begin{figure*}
    \centering
    \includegraphics[width=\textwidth]{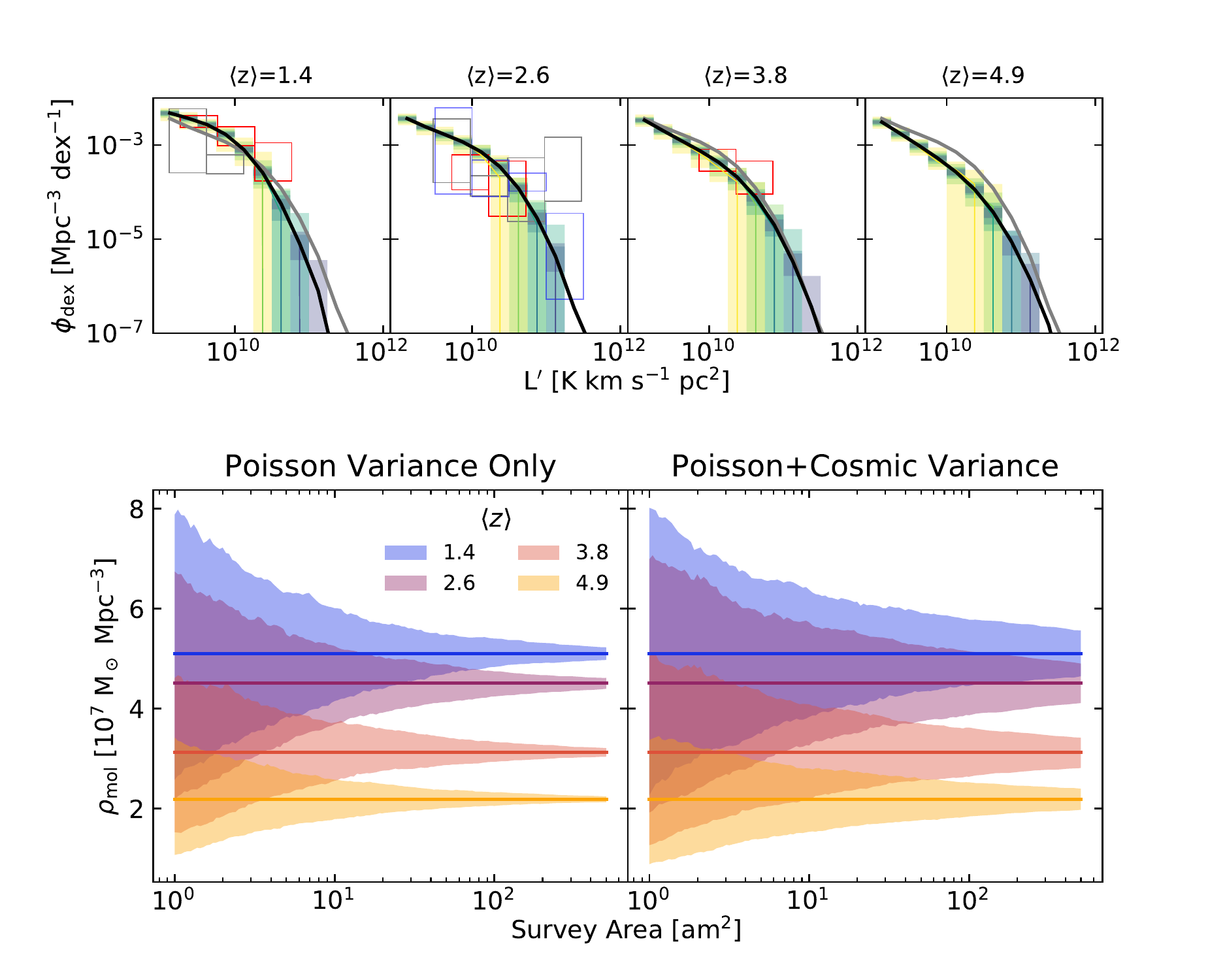}
    \caption{Top: From left to right we show the results of evaluating our modeled luminosity function at redshifts $\langle z\rangle=1.4$, $2.6$, $3.8$, and $4.9$ corresponding to the redshift windows probed by CO(2-1), (3-2), (4-3), and (5-4) respectively in ASPECS. Colored lines and filled boxes show the median and 16th to 84th percentile range of CO luminosity functions recovered by surveys of a range of sizes (colored to match Figure~\ref{fig:lf}). The black and gray lines show the true model luminosity function for the redshift window and redshift $2.6$ respectively. Plotted with open boxes are the luminosity function measurements from \citet{riechers+18} (blue), \citet{decarli+19} (red), and \citet{lenkic+20} (gray). Bottom: 16th-84th percentile range (filled regions) as a function of area and true value (horizontal lines) of $\rho_{\rm mol}$ for each redshift, converted from $\langle T\rangle$ using Equation~\ref{eq:Trho} and $\alpha_{\rm CO}=3.6$ M$_\odot$ (K km s$^{-1}$ pc$^2$)$^{-1}$. On the left side we account only for Poisson variance, while on the right side we include both Poisson and cosmic variance, resulting in considerably wider uncertainties.}
    \label{fig:rs}
\end{figure*}

Up to this point we have restricted our discussion to redshift $\langle z\rangle=2.6$. In order to assess how the luminosity function and its moments evolve, we have repeated our analysis for the redshift ranges 1.0--1.7, 3.0--4.5, and 4.0--5.9, corresponding to the redshift regimes probed in CO(2-1), CO(4-3), and CO(5-4) by the 3mm ASPECS observations. The top panel of Figure~\ref{fig:rs} shows the median and range of luminosity functions along with the observational constraints on the luminosity function from ASPECS, COLDz, and PHIBSS2 (open boxes). The ASPECS and PHIBSS2 results are scaled from higher J CO luminosities to CO(1-0) using $L_{CO(J-(J-1))}/L_{CO(1-0)}$ of $0.76$, $0.42$ and $0.31$ for $J=2$, $3$, and $4$ \citep{daddi+15,decarli+19}. We also show the true luminosity function from our model at each redshift and at redshift 2.0--3.1 with black and gray lines respectively.

Our model results in evolution of the shape and normalization of the luminosity function. There is a drop in the normalization between $\langle z\rangle=3.8$ and $\langle z\rangle=4.9$. Section~\ref{sec:direct} implies that this will make it more difficult for small surveys to measure the bright end. We indeed see that some 5 arcmin$^2$ surveys cutoff well below the knee of the luminosity function. But this effect is lessened somewhat by the larger volume sampled at higher redshifts for a survey of fixed area and frequency coverage.

Because the conversion between mean brightness temperature and mean molecular gas density (Equation~\ref{eq:Trho}) depends on redshift, constant temperature with redshift does not imply constant molecular gas density. Therefore to compare redshifts we convert our mean brightness temperatures to mean molecular gas densities using Equation~\ref{eq:Trho} and $\alpha_{\rm CO}=3.6$ M$_\odot$ (K km s$^{-1}$ pc$^2$)$^{-1}$.
The bottom panels of Figure~\ref{fig:rs} show the 16th-84th percentile range of measured mean molecular gas density as a function of area for each redshift window (filled regions), along with the true value at each redshift (horizontal lines). In the left panel we show the ranges for Poisson light cones with no large scale structure, while in the right, we show light cones with large scale structure included. The density peaks in the redshift range $1<z<2$ and falls at later times. 
From this figure we see that surveys wishing to constrain the cosmic evolution of molecular gas likely need very large survey areas (hundreds of square arcminutes) in order to reliably distinguish the evolutionary signature from Poisson and cosmic variance for redshifts $1<z<3$.

For the second moment, Poisson uncertainties are the dominant contribution to the errors for all survey sizes and redshifts. We show the evolution of the second moment in Appendix~\ref{appendix:evo2m}. Changes in the shape of the bright end of the luminosity function produce an increase in the second moment from redshift 1.4 to 2.6. As a result, for redshifts $1<z<3$ evolution in the second moment can be detected in much smaller fields than evolution in the first moment. On the other hand, evolution at higher redshifts is more difficult to detect in our models. This highlights the complimentary information that can be extracted from measurements of both moments in intensity mapping surveys.

\section{Effects of Altering Model Parameters}\label{sec:modeldependence}

\begin{figure*}
    \centering
    \includegraphics[width=\textwidth]{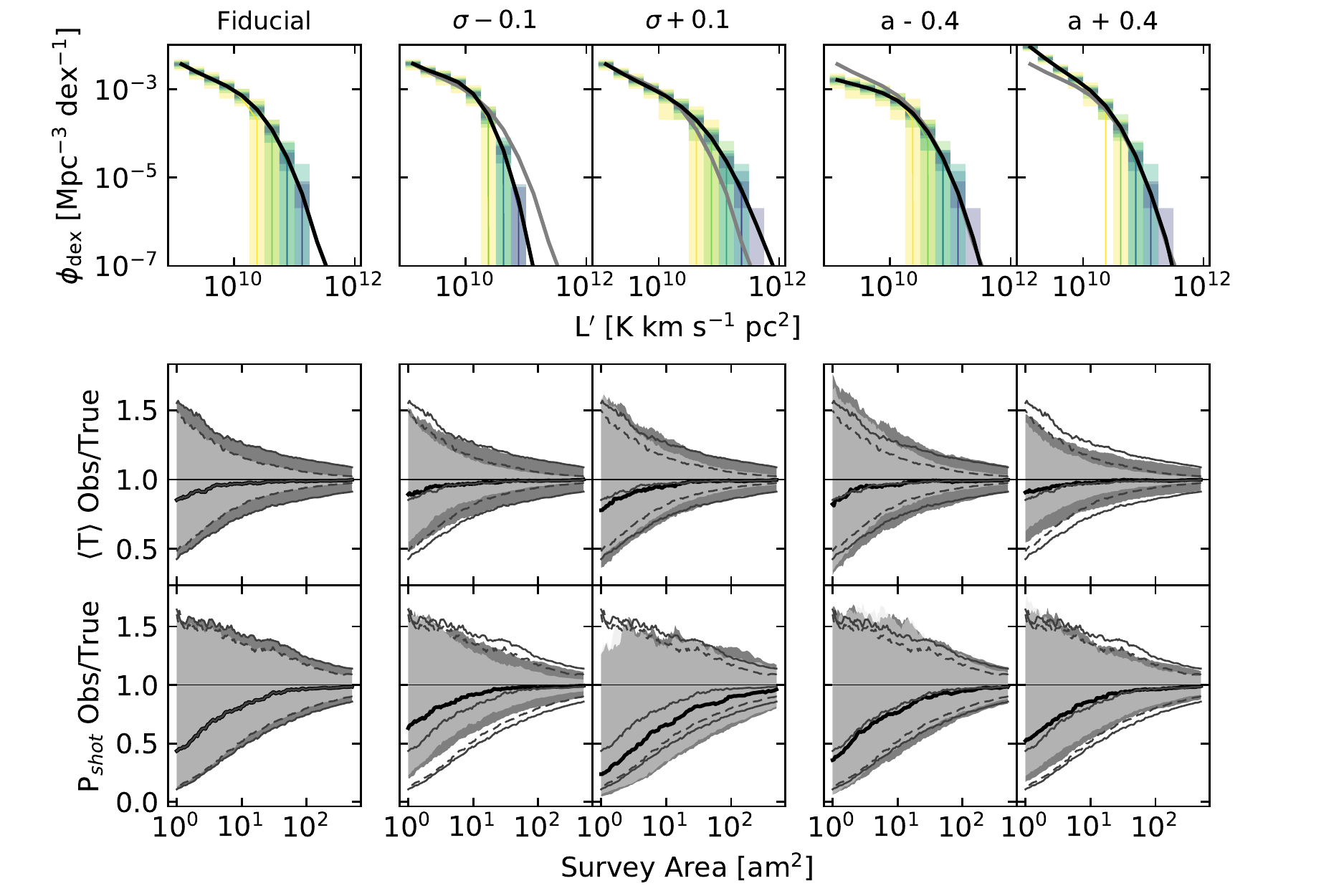}
    \caption{Results from our fiducial model (column 1) and variations on our model with the scatter parameter decreased/increased by 0.1 dex (column 2/3) or the $L_{\rm IR}$ to $L_{\rm CO}$ relation changed by decreasing/increasing $a$ by 0.4 (column 4/5). In the top row we show the median and 16th-84th percentile ranges for the luminosity functions of light cones of a range of sizes (colored to match Figure~\ref{fig:lf}), along with the true luminosity function in black and the luminosity function of the fiducial model in gray. The second row shows the median (black line) and range (filled area) of mean brightness temperatures, normalized by the true value, as a function of survey area. Dark contours are the spread of light cones with cosmic variance included, while light colors show only the Poisson contribution. The third row shows the same quantities for the shot power. We reproduce the median and range of our fiducial model with gray lines in each panel. The range of Poisson light cones for our fiducial model is shown with dashed gray lines.}
    \label{fig:params}
\end{figure*}

The exact degree and nature of many of the biases that we have identified depend on the underlying shape of the luminosity function. This is most easily seen in the left panel of Figure~\ref{fig:lfmom}. In the figure, a wide range of luminosities contribute relatively equally to the first moment. If the faint-end slope of the luminosity function were altered, the contribution of faint galaxies would change, shifting the weight given to rarer galaxies around the knee of the luminosity function. In the case of very flat slopes, this would result in a larger downward bias in the recovered mean brightness temperature for small survey areas.

To explore these effects we modified our fiducial model in two ways. First, we changed the scatter in our IR and CO luminosity prescription (keeping the mean value fixed). This tends to alter the number of objects which scatter up from low luminosity bins to higher bins, which moves the knee of the luminosity function to higher luminosities for larger scatter. We apply scatters of $\sigma=0.25$ and $0.45$ dex (0.1 dex less or more than our fiducial model).

Second, we altered the slope and y-intercept of IR to CO luminosity prescription in Equation~\ref{eq:IRCO}. \citet{li+16} find that the fits for this relation from a number of previous high redshift studies fall near 
\begin{equation}\label{eq:Lcoconst}
    a = 0.10b + 1.19
\end{equation}
and so we use $(a,b)$ pairs $(0.97,2.2)$ and $(1.77,-5.8)$ along this line. The primary effect of increasing $a$ (decreasing $b$) is to steepen the faint-end slope of the luminosity function. Note that the variations we explore for $(a,b)$ are intentionally extreme in order to illustrate that our results are insensitive to large changes in the luminosity function shape. Not all of these represent realistic models; $a=1.77$ implies a slope for the integrated Kennicutt-Schmidt relation much steeper than supported by most observations at high redshift \citep{carilli+13,greve+14,dessauges-zavadsky+15,freundlich+19,aravena+19}.

The combination of these two model variations allows us to explore a range of luminosity function shapes. Figure~\ref{fig:params} shows the luminosity functions, mean brightness temperatures, and shot powers as a function of area for our fiducial model and the full set of model variants described above. The mean luminosity function, and median and range of moments for our fiducial model are reproduced with gray lines in each column to facilitate comparison. 

Our primary findings are not affected by model choice. In all cases, surveys of very small areas fail to recover objects above $L_*$, and measure moments of the luminosity function considerably less than their true values. The degree of these biases is somewhat dependent on input model. 

Broadly, increasing the scatter produces brighter and rarer objects, reducing the likelihood of the median survey accurately measuring the bright end. As $\sigma$ increases from 0.25 to 0.45, $L_*$ grows relative to the luminosity of the brightest object likely to be recovered in small surveys, making it harder to measure the shape of the bright end. Also, the downward bias in the moment measurements becomes more pronounced, even for larger surveys. For the second moment, at $\sigma=0.45$, the median has not completely converged with the mean even for the largest survey sizes.

As \citet{li+16} point out, many of the papers from which equation~\ref{eq:Lcoconst} are derived use similar or overlapping data sets. To first order then, differing fits are just different ways of drawing a line that passes through the subset of galaxies that are both bright enough to be detected and common enough to be selected. As such the effects of changing $(a,b)$ are limited mostly to the faint end.

These changes have a less pronounced effect on the measurement of the luminosity function. The shallow faint-end slopes produced by small values of $a$ also decrease the weight given to the faint end in the moment measurements, resulting in increased fractional variance. Shallow slopes also cause a slightly greater downward bias in the moment measurements, although for the range of parameters explored here this effect is negligible.

\section{Discussion}\label{sec:discussion}
\subsection{Implications for The Shape of the CO Luminosity Function}\label{sec:directdiscussion}

ASPECS sought to determine the CO luminosity function out to $z\sim4$ and place constraints on the redshift evolution of the cosmic molecular gas density \citep{decarli+16,decarli+19}. The small sky area used for this survey inherently limits the range of luminosities that can be recovered - the galaxies which determine the bright end behavior of the luminosity function are rare, and may not reliably appear in fields the size of the HUDF. Our modeling finds that the result of this is a downward bias in the fitted knee of the luminosity function.  Over limited ranges in $L$, the parameters of the Schechter fit are highly degenerate.  Observational studies of the UV luminosity function found that decreasing $\alpha$ and increasing $\phi_*$ at fixed $L_*$ can produce comparable fits to decreasing $L_*$ \citep{bouwens+15}. This is clear from our Figure~\ref{fig:mcmccorner}, where even using galaxies down to 1.5 orders of magnitude below $L_*$, there are large degeneracies in all parameters for fields smaller than 500 arcmin$^2$ owing to incompleteness at the bright end. A similar effect seems to have been at least partially responsible for apparent evolution in the characteristic UV luminosity observed at $z>5$ by studies completed before the full CANDELS data set became available \citep{bouwens+15}.

In reality, the range in $L$ probed by ASPECS is approximately 1 dex, which will worsen these degeneracies. In fitting their luminosity functions, \citet{decarli+19} fix the value of $\alpha$ to $-1.2$ at all redshifts based on results from lower redshift \citep{saintonge+17}. 
However, studies of the UV luminosity function find redshift evolution of the faint-end slope \citep{reddy+09,bouwens+15,finkelstein+15}, and the correlation between the star formation powering the UV luminosity and the molecular gas content \citep{kennicutt+12} suggests that the CO luminosity function might evolve similarly. If there is unmodeled evolution, the parameter degeneracies will translate the error in the assumed $\alpha$ to biases in the fitted values of the other parameters. If the $z\sim2.5$ slope of the CO luminosity function is steeper than assumed by the ASPECS fits, this would have the effect of decreasing the fitted $L_*$ value, potentially resulting in an underestimate of both the bright- and faint-end number densities. 

The COLDz survey observed CO(1-0) at $z\sim2.4$ over an area $\sim 10$ times larger than ASPECS. The sensitivity and array setup of their wide field observations varied over the survey area, which changes the effective area over which COLDz is sensitive to objects of a given brightness. \citet{pavesi+18} account for sensitivity variations by injecting simulated sources and measuring the fraction recovered as a function of flux and line width. Number densities of detected sources are then corrected by this fraction. For this procedure to work, the detected objects must be representative of the unrecovered ones. For bright galaxies, this assumption fails, because there may be only a handful in the whole volume. This can cause the biases we found for small surveys to apply in surveys with larger nominal areas. Figure~21 of \citet{pavesi+18} suggests that the completeness for $L_*$ objects can fall well below 50\% in the COLDz wide field, therefore the effective area over which $L_*$ galaxies can be detected is much less than 50 arcmin$^2$ and their bright end fits may be subject to considerable bias.

\citet{lenkic+20} report the results of searching many independent data cubes from the targeted PHIBSS2 observations for additional serendipitous sources, and use these detections to constrain the CO luminosity function in a manner similar to a blind survey. The volume searched is comparable to COLDz, although spread over a wider range of redshifts. Unfortunately, the large variations in integration time between their fields makes an assessment of the volume over which different sources might be detected impossible here. We note, that their $\langle z\rangle=2.4$ luminosity function shows an excess of bright sources compared to COLDz, suggesting the volume covered by these two surveys is not yet large enough to fully constrain the knee of the luminosity function.

When computing the UV luminosity function, \citet{finkelstein+15} convert their completeness corrections into effective volumes which they report as a function of luminosity. Taking a similar approach may be useful in future CO line scan results, and can be used to more fairly represent the true volumes probed when comparing surveys.

The wide-field extension to ASPECS will search for CO emitters over an area of $\sim 50$ arcmin$^2$ in a more limited redshift range and may begin to improve the situation described here, assuming that it can reach the needed depth to detect objects around $L_*$ with more uniform coverage than COLDz. Experience from luminosity function studies at other wavelengths suggests that surveys of hundreds of arcmin$^2$ will likely be necessary to fully constrain the shape of the luminosity function.

\subsection{Implications for Cosmic Molecular Gas Density}\label{sec:rhodiscussion}
\subsubsection{Sensitivity Limitations}

The results of Section~\ref{sec:direct} also have implications for efforts to measure the cosmic molecular gas density from direct measurements. Current-generation surveys likely cannot constrain the shape of the luminosity function with high precision. The parameter degeneracies and lack of dynamic range in fitting the shape of the luminosity function make it impossible to draw conclusions about the relative importance of faint galaxies. The COLDz fit for the faint-end slope allows a range of $\alpha=0.0$ to $\alpha=-1.8$ (not meaningfully different from their uniform prior), and the PHIBSS2 fit at lower redshift ($\langle z\rangle=0.7$) also allow a wide range. Without better constraints on this parameter, line scans cannot independently constrain the contribution of faint galaxies to the total molecular gas density.

To avoid extrapolation of the LF, recent blank field searches have assessed the molecular gas density from their detections alone. Figure~\ref{fig:cut} shows that this approach can recover the true mean brightness temperature, and therefore $\rho_{\rm mol}$ contribution, of these galaxies with relatively little bias. On the other hand, the fraction of the total molecular gas density recovered is uncertain. In addition, since the luminosity range recovered varies with redshift, directly comparing the molecular gas density at different redshifts can produce apparent evolution purely due to selection effects.

\citet{popping+19} also explore biases the ASPECS measurement of $\rho_{\rm mol}$ in the context of comparisons with more sophisticated simulations of the molecular gas content of galaxies. Their approach simulates a molecular gas mass which must then be converted to CO luminosity by assuming a value of $\alpha_{\rm CO}$. We show in Figure~\ref{fig:lfmodel} the luminosity functions produced by the semi-analytic model of \citet{popping+19} adopting both $\alpha_{\rm CO}=3.6$ M$_\odot$ (K km s$^{-1}$ pc$^2$)$^{-1}$, found by \citet{daddi+10} to be apropriate for redshift 1.5 star forming galaxies, and $\alpha_{\rm CO}=0.8$ M$_\odot$ (K km s$^{-1}$ pc$^2$)$^{-1}$, appropriate for local starburst galaxies \citep{downes+98}. Both models struggle to reproduce the full shape of the luminosity function implied by observational data. The $\alpha_{\rm CO}=3.6$ M$_\odot$ (K km s$^{-1}$ pc$^2$)$^{-1}$ luminosity function drops off sharply around the sensitivity limit of ASPECS and produces too few bright objects. A choice of $\alpha_{\rm CO}=0.8$ M$_\odot$ (K km s$^{-1}$ pc$^2$)$^{-1}$ gives a higher recovered fraction, but still does not match the shape of the CO luminosity instead producing too many faint galaxies.

Observational constraints on $\alpha_{\rm CO}$ from redshifts 1.0--1.6 favor a Milky Way-like value \citep{carleton+17,daddi+10}. The models in \citet{popping+19} would then suggest that a survey with ASPECS-like area and depth only recovers $\sim10$\% of the total molecular gas mass density at $z>2$. On the other hand, our Figure~\ref{fig:cut} suggests that a survey with a cutoff of $\log L^\prime_{\rm CO(1-0)}=9.45$ (approximately the sensitivity reached by ASPECS and comparable to the threshold assumed in \citealt{popping+19}) recovers $77_{-27}^{+34}$\% of the CO luminosity galaxies with $L^{\prime}>10^9$ K km s$^{-1}$ pc$^2$. Assuming fainter galaxies account for $\sim 25$ percent of the total luminosity (see Section~\ref{sec:estmoments}) gives an overall recovery rate of $55_{-19}^{+24}$\%. In contrast to the results of \citet{popping+19} this suggests that ASPECS is accounting for a sizable fraction of the total molecular gas density at least out to $z\sim3$, without the need to invoke small values of $\alpha_{\rm CO}$.

As \citet{popping+19} point out, there are a number of unresolved issues in galaxy theory that make matching observations with physics based models challenging. The purpose of our models is to produce realistic forecasts using empirical relations and scaling laws, which allows us to circumvent many of these issues. As our model reasonably reproduces observations over the full luminosity and redshift range considered, it seems plausible that current surveys do recover a substantial fraction of the total molecular gas density.

Intensity mapping experiments directly constrain the integral of the luminosity function over all luminosities, and can recover the mean brightness temperature with no luminosity cutoff. This will allow them to provide a check on the cutoff related biases identified here and in \citet{popping+19}. In addition to filling in missing information about the faint end, upcoming intensity mapping experiments will survey areas of 100 to 1000s of arcmin$^2$, potentially making them more robust to field to field variations and downward biases in luminosity function moments.

There are a number of complications in constraining $\rho_{\rm mol}$ via intensity mapping. The CO clustering power spectrum, from which the mean brightness temperature is derived, also depends on the tracer bias $b_{\rm CO}$ and the matter power spectrum $P_{\rm lin}$, which introduce their own uncertainties. In addition, the CO luminosity to molecular gas mass conversion factor $\alpha_{\rm CO}$ varies considerably among different galaxy populations. If this results in variations in the mean $\alpha_{\rm CO}$ over different parts of the luminosity function, then a simple scaling of the first moment of the luminosity function does not recover $\rho_{\rm mol}$. Moreover, intensity mapping experiments using higher excitation transitions of the CO lines and must invoke a conversion, $r_{J1}$, to convert to CO(1-0) luminosity. This conversion factor varies from galaxy to galaxy depending on the conditions of the ISM, and has only been measured for a handful of ``typical'' star forming galaxies at high redshift. In current results from both intensity mapping and direct detection, both $\alpha_{\rm CO}$ and $r_{J1}$ are assumed to be known constants. In principle, direct detection surveys provide catalogs of galaxies that can be followed up to produce improved constraints on these quantities for each object being considered. Intensity mapping does not provide targets for directed follow up. 

As this field develops, joint fitting of intensity mapping and direct detection constraints may provide insight into the shape of the luminosity function well below the detection limit. Combined with trends in $\alpha_{\rm CO}$ with $L^\prime$ or other galaxy properties identified by direct detection surveys this would allow for computation of $\rho_{\rm mol}$ by integrating over the full luminosity function.
    
\subsubsection{The Importance of Cosmic Variance}

Figure~\ref{fig:TvA} shows that the variance in the mean brightness temperature and $\rho_{\rm mol}$ exceeds the expectation for Poisson statistics at all survey sizes. For an ASPECS sized survey, cosmic variance widens the 16th-84th percentile range by about 30\%, and for a survey the size of COLDz it fully doubles the range. Treatments of cosmic variance in observational papers about the CO luminosity function have tended to dismiss it as a secondary effect and not included it in reported errors. Our results suggest that this results in a substantial underestimation of the uncertainty in $\rho_{\rm mol}$.

\citet{riechers+18} and \citet{lenkic+20} report estimates of cosmic variance using the formula from \citet{driver+10}. That work considered the variance in galaxy counts based on samples of $z=0.03-0.1$ $M^*_R$ galaxies, though not the variance of luminosity moments of those counts, and intentionally sought to avoid effects from highly biased galaxies that might dominate surveys of the very brightest CO line emitters. To provide a more apposite prediction, in Appendix~\ref{sec:prescription} we provide a prescription for estimating cosmic variance in $\rho_{\rm mol}$. Comparison between our prescription and prior results suggests that care must be taken in identifying applicable methods for determining sample variance. The quantities being measured as well as the clustering of the galaxies under consideration can significantly alter estimates of sample variance, by up to a factor of four in some cases.

The cosmic molecular gas density is expected to evolve in a manner analogous to the cosmic star formation rate density \citep{tacconi+18}. Testing this expectation is a central goal of studies of molecular gas at high redshift. Figure~\ref{fig:rs} shows that when cosmic variance is accounted for, definitively identifying evolution in $\rho_{\rm mol}$ becomes much more difficult than has been appreciated. For surveys of a single contiguous field, we find that even a 500 arcmin$^2$ area (an order of magnitude larger than the largest direct detection surveys completed thus far) cannot detect the evolution in our model from $z\sim1$ to $3$ during the peak of cosmic star formation.

The exact evolution is model dependent, but Figure~\ref{fig:rs} gives a sense for the degrees of evolution that are identifiable in different survey sizes. The star formation rate density rises and then falls by a factor of $\sim2$ and from redshifts 4 to 2 \citep{madau+14}. If the SFRD and molecular gas histories are comparable, then comparing redshifts 1.4 and 3.8 in Figure~\ref{fig:rs} suggests factors of $\sim2$ are indistinguishable in a 5 arcmin$^2$ survey and only reach a significance of $\sim 2\sigma$ in a 50 arcmin$^2$ survey. This implies that identifying this degree of evolution with high confidence will require surveys of hundreds of arcmin$^2$, especially if narrower redshift intervals are to be considered. Current generation surveys will therefore have difficulty constraining models of galaxy evolution during the epoch of galaxy assembly or locating the precise peak of the molecular gas density history. Over a wider redshift range, the evolution is likely to be more pronounced (e.g. SFRD rises by a factor of $\sim 10$ from redshift $\sim6$ to $2$), and should be detectable in relatively small fields if surveys can reach the required sensitivity to study $z\sim5$ objects.

On the other hand, Figure~\ref{fig:rs} also suggests that if cosmic variance can be mitigated the total survey area required to detect evolution in $\rho_{\rm mol}$ falls substantially. This can be achieved by combining multiple small fields in widely separated parts of the sky. For identical fields, uncertainty due to cosmic variance falls as $1/\sqrt{N_{\rm fields}}$ where $N_{\rm fields}$ is the number of independent fields \citep{moster+11,driver+10}. A 50 arcmin$^2$ survey composed of $10 \times 5$ arcmin$^2$ fields will therefore have $\sim3$ times less cosmic variance than a contiguous field. Cosmic variance has the greatest effect for objects at the faint end of the luminosity function, therefore this approach may be particularly beneficial for ``deep'' fields aimed at detecting the faintest objects.

Cosmic variance may present a particular challenge for intensity mapping studies, where achieving the needed sensitivity to the large scale modes of the power spectrum containing information about the first moment requires surveys over large \textit{contiguous} volumes. This requirement will limit the number of independent volumes upcoming intensity mapping surveys can study. Figure~\ref{fig:TvA} demonstrates that as these surveys probe larger areas, cosmic variance greatly increases the total sample variance over the Poisson only case. Clustering power measurements will have to carefully account for this uncertainty in their analyses. Equation~\ref{eq:ps} shows that field to field offsets in the mean CO brightness temperature will produce real shifts in the normalization of the clustering power term of the CO power spectrum. This is distinct from the case of the matter power spectrum, where normalization of the density fluctuations by mean density should eliminate these offsets and sampling uncertainties will fall in proportion to the number of modes squared.

\subsection{Implications for Shot Power Intensity Mapping}\label{sec:imdiscussion}

The first generation of CO intensity mapping experiments lacked the volume and sensitivity to measure the (large-scale) clustering power component of the CO power spectrum. Instead they focused on measuring the shot power. The shot power is proportional to the second moment of the luminosity function, with no other parameter dependencies, giving it the added benefit of being simple to interpret. In addition, it can be measured in smaller volumes, making it possible to reduce cosmic variance by targeting numerous moderate size fields. Figure~\ref{fig:lfmom} shows that the first and second moment are primarily determined by different portions of the luminosity function, with the second moment being most sensitive around the turnover.

Because both the shot power measurement and current direct detection efforts sample around the knee of the luminosity function, the second moment is also where existing intensity mapping and direct detection efforts can most easily be compared. However, care must be taken when making these comparisons. While small fields produce relatively unbiased measurements of the first moment of the luminosity function, Figure~\ref{fig:SvA} shows that the median shot power recovered in an ASPECS sized field is significantly lower than the true value.

COPSS II measured a power of $3000\pm1300$ $\mu$K$^2$ $h^{-3}$Mpc$^3$, with most of their sensitivity coming from the shot power regime \citep{keating+16}. Based on a revised analysis of the COPSS II data \citet{keating+20arxiv} report that the shot power accounts for $2000^{+1100}_{-1200}$ $\mu$K$^2$ $h^{-3}$Mpc$^3$ of the total. Section~\ref{sec:cvmoments} suggests that cosmic variance has little effect on the shot power for individual COPSS II fields, which have sizes of $\sim140$ arcmin$^2$. Combining multiple, widely spaced fields helps to further minimize cosmic variance. We simulate the results of a COPSS II-like survey by averaging the shot power for sets of 17 light cones, using the same weighting scheme employed by \citet{keating+16} and find a fractional uncertainty of 18\%, in good agreement with their reported estimate.

\citet{uzgil+19} used the ASPECS direct detections to estimate the CO shot power\footnote{They also use the fitted Schechter functions in a similar analysis, but for reasons discussed in Section~\ref{sec:directdiscussion}, we find lower limits from direct detections to be more reliable for comparison}. They find that the direct detections in the ASPECS field imply a lower limit to the shot power of 98 $\mu$K$^2$ $h^{-3}$Mpc$^3$ for CO(3-2) at $\langle z \rangle = 2.6$. To compare this to the COPSS II result, we convert the CO(3-2) result to CO(1-0) using a $L^\prime_{\rm CO(3-2)}/L^\prime_{\rm CO(1-0)}=0.42$, resulting in a minimum shot power from the ASPECS field of 560 $\mu$K$^2$ $h^{-3}$Mpc$^3$. \citet{uzgil+19} point out that these limits are lower than the COPSS II central value. However, the fractional uncertainty in the shot power for a 5 arcmin$^2$ field is around 80\%, as shown in Figure~\ref{fig:SvA}. Furthermore, the median recovered shot power is only $\sim70$\% of the true value and a full two thirds of our simulated fields recover less than the true value. These effects can account for much of the discrepancy between the two results.

The first phase of mmIME produced constraints on the CO power spectrum at 3mm over roughly quadruple the area of ASPECS. \citet{keating+20arxiv} use the simulations presented here to correct for downward biases in the recovered shot power. They find a CO(3-2) shot power of $210^{+110}_{-80}$ $\mu$K$^2$ $h^{-3}$Mpc$^3$ after using a model prescription to divide their measured power between CO transitions. Converting to CO(1-0) this gives $1190^{+900}_{-670}$ $\mu$K$^2$ $h^{-3}$Mpc$^3$, which is consistent with the COPSS results. By contrast, an analysis using only instrumental noise for uncertainties and not accounting for the biases discussed here would have given a CO(3-2) shot power of $160\pm40$ $\mu$K$^2$ $h^{-3}$Mpc$^3$. This demonstrates that by failing to account for sampling effects surveys may report parameter estimates offset from the true mean value. Further, they can considerably underestimate the size of their errors, even when instrumental noise or other sources of uncertainty are large.

\subsection{Implications for Dust Continuum Studies of Molecular Gas}\label{sec:dustdiscussion}

In recent years the use of dust continuum photometry has been established as an alternative method of measuring molecular gas masses. In this method, measurements at wavelengths in the Rayleigh-Jeans tail of galaxies' dust emission are used to determine total gas masses using empirical calibrations of the ratio of dust continuum and CO(1-0) luminosity \citep[e.g.][]{scoville+17,liu+19,magnelli+20}. This method has gained increasing attention because the dust continuum emission can be detected at high redshift with comparatively short integration times.

A full consideration of the details of this method are beyond the scope of this paper. We only note that, as the empirical basis of this method is a straightforward relation between CO and dust emission, our results should be applicable to dust-based as well as CO-line-based observations. The same bright CO emitters that are missed by small area surveys will be absent from dust-based molecular gas censuses over similar areas, producing similar biases in deep field searches for dust emission to those reported for CO line scans. 

The reduced integration times for dust continuum measurements relative to CO lines means that it may be possible to probe the luminosity function of fainter galaxies with this method. However, this method has so far only been calibrated for fairly bright samples \citep{liu+19}, and so may not produce results consistent with the CO approach for low luminosity objects. Different systematic uncertainties of these two methods may also make synthesis difficult.

\section{Conclusion}\label{sec:conclusion}

We have used simulated light cones populated with CO emitting galaxies to explore some of the challenges facing current and upcoming efforts to measure the shape of the CO luminosity function and the evolution of the cosmic molecular gas density. We find that there are a number of potential biases due to the small volumes to which these surveys are currently restricted. These will hinder interpretation of results if not carefully accounted for.

Our primary conclusions are as follows:
\begin{enumerate}
    \item The properties of the bright end of the CO luminosity function are determined by rare objects which may not reliably appear in survey fields with areas of 50 arcmin$^2$ or less. As a result, fits to the measured luminosity function may suffer from significant defects, where $L_{*}$ and $\phi_{*}$ are offset from their true values. The exact field size required to achieve accurate constraints will depend on the true luminosity function and cannot be determined in advance.
    \item Cosmic variance, which has generally been dismissed as subdominant to other sources of error for current surveys, can have a significant effect on some, though not all, quantities measured. The apparent shape of the luminosity function is mostly unaltered by cosmic variance for current surveys, but the normalization, when expressed in terms of mean number density of CO emitters shows considerably larger uncertainty, even in surveys comparable in size to ASPECS.
    \item The first moment of the luminosity function, which is proportional to the cosmic molecular gas density, is determined by contributions from galaxies at a wide range of luminosities. Measurements of this quantity are relatively unbiased by small survey sizes, and surveys sensitive to galaxies below the knee of the luminosity function likely recover a substantial fraction of the total molecular gas density.
    \item However, cosmic variance contributes appreciably to the uncertainty in the mean brightness temperature and cosmic molecular gas density for surveys of all sizes. For surveys larger than $\sim 50$ arcmin$^2$, it may be the dominant source of error. The volume required to detect evolution around the peak of cosmic star formation increases by and order of magnitude, from tens of square arcminutes to hundreds, when this is accounted for. This uncertainty must also be accounted for in intensity mapping analyses of the clustering term in the CO power spectrum.
    \item Appendix~\ref{sec:prescription} provides a means to estimate the fractional sample uncertainty, including cosmic variance, in the mean brightness temperature and shot power as a function of redshift which can be scaled to different survey areas and redshift intervals.
    \item Surveys divided over multiple small fields mitigate cosmic variance and can reduce the total field size required to detect redshift evolution of the cosmic molecular gas density to a scale achievable with current generation instruments
    \item The second moment of the luminosity function, which has been constrained by a handful of intensity mapping experiments, is subject to larger biases when measured in small fields. Using surveys like ASPECS and COLDz-COSMOS to forecast this quantity will likely result in underestimates by as much as 30\%. The apparent tension between current direct detection and intensity mapping results found by \citet{uzgil+19} is significantly reduced when this bias is accounted for.
\end{enumerate}

Recent results on the CO luminosity function and $\rho_{\rm mol}$ have estimated error bars using only Poisson uncertainties on number counts. This practice fails to represent the full set of uncertainties faced by these measurements.

Taken together, our results suggest that definitive measurement of the CO luminosity function and cosmic molecular gas density will require larger surveys than have currently been undertaken. The combination of intensity mapping and direct detection surveys may provide a promising path forward. Ongoing and planned intensity mapping experiments such as the CO Mapping Array Pathfinder \citep[COMAP;][]{li+16}, the Tomographic Ionized-carbon Mapping Experiment \citep[TIME;][Sun et al. in prep]{crites+14}, and the CarbON CII line in post-rEionization andReionizaTiOn epoch project \citep[CONCERTO;][]{dumitru+19} will place integral constraints on the CO luminosity function over a broad redshift range using surveys areas ranging from hundreds of square arcminutes to more than a square degree. Meanwhile, CO emitting galaxies identified by direct detection can be characterized in detail providing needed insight into how the molecular gas properties such as the CO luminosity to molecular gas mass conversion factor and the CO line excitation ratios vary as a function of CO luminosity and other galaxy properties. These detailed snapshots can be used to develop more sophisticated interpretations of the large area constraints provided by intensity mapping surveys, making it possible to construct improved estimates of the CO luminosity function and cosmic molecular gas density.

\acknowledgments
We would like to thank the anonymous referee for their thorough and constructive feedback. We would also like to thank P. Behroozi for insightful discussions throughout the development of this research, X. Fan for guidance with fitting luminosity functions, and G. Popping for insights regarding his theoretical models. RPK was supported by the National Science Foundation through Graduate Research Fellowship grant DGE-1746060. DPM and RPK were supported by the National Science Foundation through CAREER grant AST-1653228.

\appendix
\section{A Prescription for Sample Variance}\label{sec:prescription}

In this Appendix, we provide a general prescription for approximating the sample variance for fields of arbitrary redshift and area. A number of works have provided prescriptions for calculating the variance in number counts, applicable to understanding the uncertainty in luminosity functions \citep{driver+10,moster+11}. However, the sample variance for moments of the luminosity function behaves differently from number counts and requires separate calculations.

\subsection{Comparison to Existing Prescriptions}

Two prescriptions for calculating sample variance have been cited frequently in literature about molecular gas density at high redshift. \citet{driver+10} calculated sample variance in number counts for ``common'' $M_R^*$ galaxies at $z\sim0$ and extrapolated these results to higher redshift. As pointed out in that paper, this formula may not apply at $z>1$ or for objects dissimilar to the $M^*_R$ galaxies studied, because the tracer bias changes with redshift and galaxy type. 
\citet{moster+11} used cosmological simulations to compute the cosmic variance in number counts (Poisson variance was not included). They present results as a function of host halo mass in addition to field geometry and redshift.

\begin{figure}
    \centering
    \includegraphics[width=\textwidth]{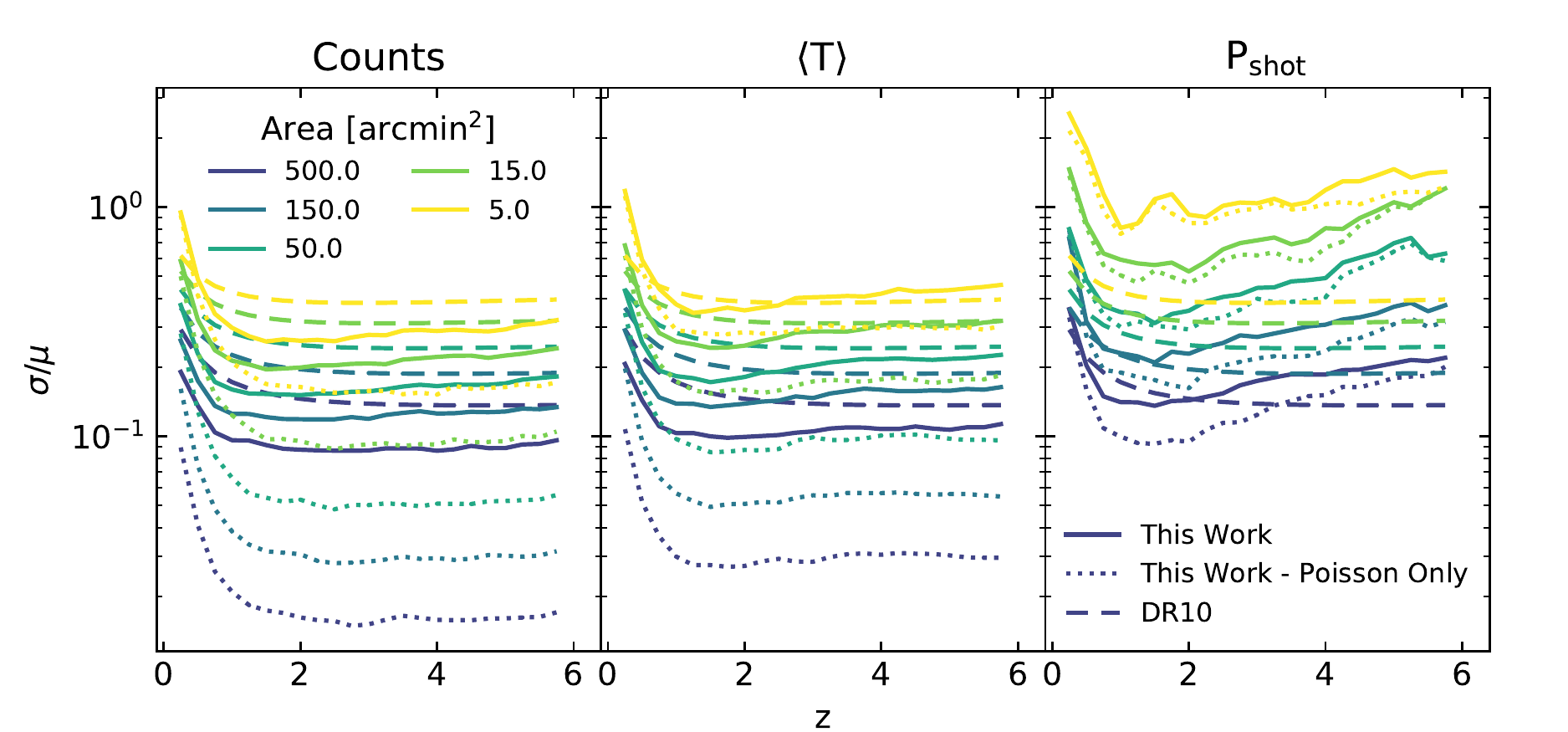}
    \caption{Standard deviation in the number of galaxies brighter than $L^\prime=10^9$ K km s$^{-1}$ pc$^2$ (left), the mean brightness temperature (center), and the shot power (right) over the mean value as a function of redshift in bins of $\Delta z/(1+z)=0.25$. Different colors show different survey sizes, and solid and dotted lines show results for light cones with and without large scale structure respectively. The dashed lines show the prescription of \citet{driver+10}, which is based on number counts for $M_R^*$ galaxies at redshift 0, but has been applied to other redshifts and quantities like mean brightness temperature.}
    \label{fig:sigrs}
\end{figure}

Both sets of results are for number counts, not moments, making their use to estimate the variance in mean brightness temperature or $\rho_{\rm mol}$ inappropriate. Our approach, of explicitly adding a CO luminosity to every halo allows us to measure sample variance in moments of the luminosity function in a way that number-count-based approaches do not. In Figure~\ref{fig:sigrs} we show the standard deviation as a fraction of the mean value for total number counts (over $L^\prime=10^9$ K km s$^{-1}$ pc$^2$),\footnote{Note that the variance for number counts depends heavily on the chosen luminosity cutoff and our results for this quantity are not meant to serve as general prescription. Instead we provide them to highlight the differences in sample variance between number counts (for which most prescriptions are calculated) and mean brightness temperature.} mean brightness temperature, and shot power as a function of redshift and survey area, computed in redshift bins of $\Delta z/(1+z)=0.25$. Comparing number counts to the moments of the luminosity function highlights that the fractional sample variance differs considerably between the different quantities. 

The dotted lines in Figure~\ref{fig:sigrs} show results for our Poisson only light cones. At all redshifts and field sizes the uncertainty in counts and mean brightness temperatures increase appreciably due to cosmic variance, whereas the shot power is Poisson variance dominated.

We also show in Figure~\ref{fig:sigrs} the predictions of \citet{driver+10} for the same field sizes, repeated in each panel. For number counts, where our results and theirs can be directly compared, our modeling gives lower fractional variance. This suggests that the tracer bias of the CO emitting galaxies in our simulations is smaller than that of the $M_R^*$ galaxies selected by \citet{driver+10}. Interestingly, when we compare our estimates fractional variance in mean brightness temperature to their number count results, we find similar values for small fields, which suggests that the use of this prescription to assess cosmic variance in $\rho_{\rm mol}$ by \citet{riechers+18} and \citet{lenkic+20} gives answers approximately in line with our results. We emphasize that this is a coincidence -- the \citet{driver+10} estimates were derived for a different quantity and for tracers with a different bias.

The right panel of Figure~\ref{fig:sigrs} shows that using results calibrated for number counts to estimate uncertainty in shot power measurements is especially misleading. \citet{uzgil+19} use the \citet{moster+11} results to estimate the fractional error due to cosmic variance in ASPECS to be 23\% at redshift 2.5. \citet{moster+11} estimate only cosmic variance, while the sample variance in shot power is primarily driven by Poisson variance. As a result, while this value is comparable to our estimate for uncertainty in number counts for ASPECS, it underestimates the total sample variance in shot power by roughly a factor of four. This highlights the importance of providing a prescription for sample variance that is tailored to the types of galaxies under study and the particular quantities being measured.

\subsection{Sample Variance as a Function of Redshift and Survey Area}

\begin{deluxetable}{cc|cccccccccccc}
\tablecaption{Fractional sample variance in a field of 50 arcmin$^2$ and $\Delta z/(1+z)=0.25$ for mean brightness temperature and shot power.\label{tab:redshift}}
\tablehead{
    \colhead{} & \colhead{} & \colhead{$z=0.5$} & \colhead{1.0} & \colhead{1.5} & \colhead{2.0} & \colhead{2.5} & \colhead{3.0} & \colhead{3.5} & \colhead{4.0} & \colhead{4.5} & \colhead{5.0} & \colhead{5.5}}
\startdata
\multirow{2}{*}{{\LARGE $\frac{\sigma}{\mu}$}}&$\langle$T$\rangle$& 0.26 & 0.18 & 0.17 & 0.18 & 0.19 & 0.20 & 0.21 & 0.22 & 0.22 & 0.22 & 0.22 \\
&P$_{\rm shot}$& 0.48 & 0.35 & 0.31 & 0.35 & 0.41 & 0.45 & 0.47 & 0.49 & 0.60 & 0.70 & 0.61 \\
\enddata
\end{deluxetable}

Table~\ref{tab:redshift} gives the standard deviation as a fraction of the mean value for mean brightness temperature and shot power for a 50 arcmin$^2$ survey with redshift interval $\Delta z/(1+z)=0.25$ for a range of redshifts. These values include the effects of both Poisson and cosmic variance.

To estimate the sample variance in fields of other sizes and redshift intervals, the values given in Table~\ref{tab:redshift} can be re-scaled. We expect that altering the survey area will result in a scaling of $\sigma\propto A^{-m}$ where $A$ is the survey area. When sample variance is primarily Poissonian, we expect $m\sim0.5$. Cosmic variance will cause the sample variance to fall more slowly with increasing area, leading to $0.0<m<0.5$. We fit for $m$ using fields from 1.5 to 500 arcmin$^2$ and find that $m$ is approximately constant for each moment in the redshift range from $1$ to $6$. Over this range we find $m=0.30$ for mean brightness temperature and $m=0.41$ for shot power. The increasing values of $m$ for higher moments reflect that greater importance of Poisson relative to cosmic variance.

So long as the transverse dimension is larger than a few hundred comoving megaparsecs, the variance should scale in a Poissonian manner when changing the redshift interval \citep{driver+10,moster+11}. We confirm this by varying redshift intervals from $\Delta z=0.05/(1+z)$ to $\Delta z=0.40/(1+z)$ and find that indeed $\sigma\propto \Delta z^{-0.5}$.

Combining these scalings, the fractional uncertainty $\sigma/\mu$ for an arbitrary survey area and redshift range can be found using
\begin{equation}\label{eq:prescT}
    \frac{\sigma_{\langle T\rangle}/\mu_{\langle T\rangle}}{(\sigma_{\langle T\rangle}/\mu_{\langle T\rangle})_{\rm ref}} = \Big(\frac{A}{50 {\rm am^2}}\Big)^{-0.30} \Big(\frac{\Delta z}{0.25(1+z)}\Big)^{-0.5}
\end{equation}
and 
\begin{equation}\label{eq:prescP}
    \frac{\sigma_{P_{\rm shot}}/\mu_{P_{\rm shot}}}{(\sigma_{P_{\rm shot}}/\mu_{P_{\rm shot}})_{\rm ref}} = \Big(\frac{A}{50 {\rm am^2}}\Big)^{-0.41} \Big(\frac{\Delta z}{0.25(1+z)}\Big)^{-0.5}
\end{equation}
where $(\sigma/\mu)_{\rm ref}$ is the the value from Table~\ref{tab:redshift}. 
Note that the appropriate area is the contiguous field area, not the total area of a survey spread among many fields. In the case of fields that are separated on the sky by at least hundreds of comoving megaparsecs, the fractional uncertainty per field can be derived from the equation above and the fields can be treated as independent samples to derive a total uncertainty. Finally, the sky-plane geometry of each field affects the variance from large scale structure, as described in \citet{moster+11} and \citet{driver+10}. The numbers above were derived for square footprints, small corrections for field geometry could be necessary in cases of extremely anisotropic footprints.

\subsection{Accounting for Survey Sensitivity}

The above fits were performed assuming our fiducial luminosity cutoff of $L^\prime=10^9$ K km s$^{-1}$ pc$^2$. We investigated how changing this cutoff alters the fractional uncertainties reported above by considering cutoffs of $L^\prime=10^{7.5}$, $10^{9.5}$, $10^{10.0}$, and $10^{10.5}$ K km s$^{-1}$ pc$^2$. Fractional uncertainties in the mean brightness temperature show little change for cutoffs from $10^{7.5}$ to $10^{10.0}$ K km s$^{-1}$ pc$^2$ in fields of 15 arcmin$^2$ or larger (shifting by no more than 25\% of the fractional uncertainty computed at $10^9$ K km s$^{-1}$ pc$^2$). For smaller fields the fractional uncertainties do not change much for cutoffs from $10^{9.0}$ to $10^{10.0}$ K km s$^{-1}$ pc$^2$ but grow for the $10^{7.5}$ K km s$^{-1}$ pc$^2$ cutoff. 

For the shot power, fractional uncertainties stay within the range of 0.75 to 1.25 times their value at $10^9$ K km s$^{-1}$ pc$^2$ for cutoffs from $10^{7.5}$ to $L^\prime=10^{9.5}$ K km s$^{-1}$ pc$^2$ before growing considerably for higher cutoffs.

This means Equation~\ref{eq:prescT} should apply to any direct detection survey sensitive down to $L^\prime=10^{10}$ K km s$^{-1}$ pc$^2$. This includes all of the surveys described in Table~\ref{tab:surveys}, except for the wide field of COLDz, which has a sensitivity limit of $L^\prime\sim10^{10.5}$ K km s$^{-1}$ pc$^2$. For this field, our simulations imply a sample variance 1.6 times higher than given by Equation~\ref{eq:prescT}. 

Equations~\ref{eq:prescT} and~\ref{eq:prescP} should be applicable to all intensity mapping experiments, which have no luminosity cutoff and generally survey areas larger than 15 arcmin$^2$. However, the sample uncertainty for using direct detection results to set limits on the shot power will be larger than implied by Equation~\ref{eq:prescP} when the direct detection survey has a sensitivity limit above $\sim10^{9.5}$ K km s$^{-1}$ pc$^2$.

\subsection{Sample Variance for Existing Surveys}

\begin{figure}
    \centering
    \includegraphics{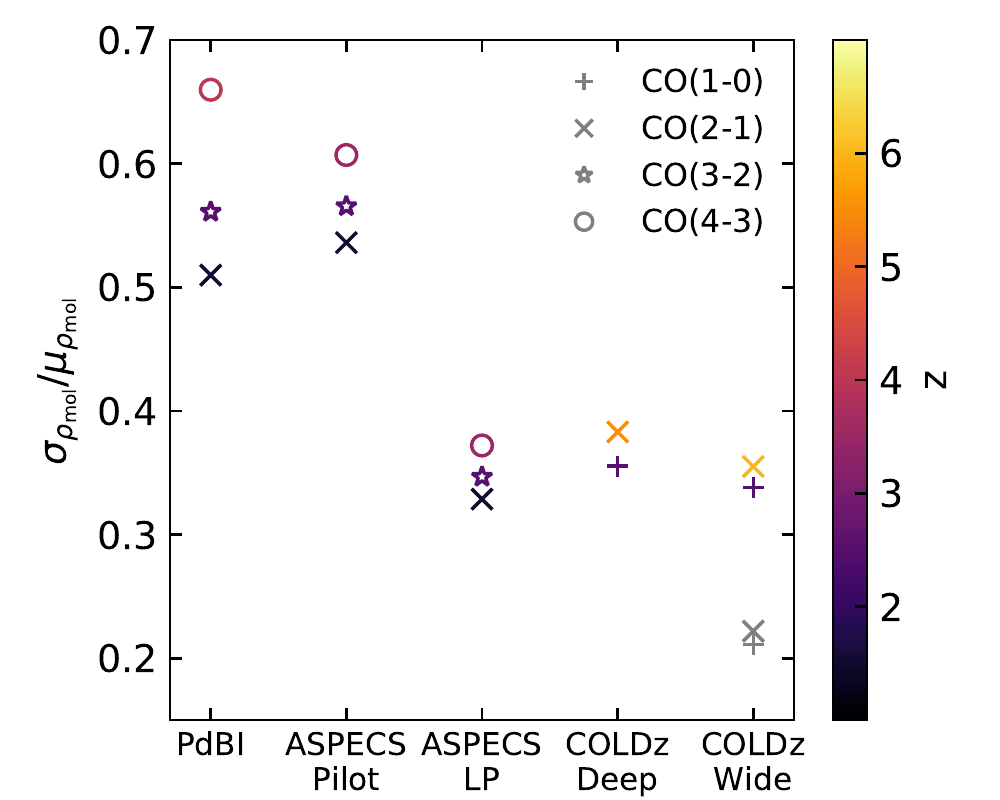}
    \caption{We show the fractional uncertainty in recovered $\rho_{\rm mol}$ estimated using Equation~\ref{eq:prescT} for a number of direct detection surveys that have been completed to date. Marker color indicates the median redshift of the survey, while marker style designated the CO line used. For the COLDz wide field, we scale up the values from Equation~\ref{eq:prescT} by a factor of 1.6 to account for shallow survey depth. We show the unscaled values for this field gray.}
    \label{fig:fields}
\end{figure}

In Figure~\ref{fig:fields} we use Equation~\ref{eq:prescT} to calculate the fractional sample uncertainties in $\rho_{\rm mol}$ for the direct detection surveys listed in Table~\ref{tab:surveys}. These uncertainties are for the density of molecular gas only from galaxies bright enough to be detected in each survey, and do not include corrections for fainter objects. We find that the most recent ASPECS and COLDz surveys all achieve fractional sample uncertainties in the 30-40\% range. For the COLDz wide field, Equation~\ref{eq:prescT} implies an uncertainty closer to 20\% (shown with gray markers in Figure~\ref{fig:fields}, however the sensitivity limits in this field increase the uncertainties to be more in line with ASPECS LP and the deeper COLDz field.

\section{Parameterizations of the Schechter Function}\label{appendix:schechter}

Here we have used the Schechter function in the form (Equation~\ref{eq:schechter})
\begin{equation}\label{eq:Spar}
    \frac{dn}{dL} = 
    \phi(L)dL = 
    \phi_* \Big(\frac{L}{L_*}\Big)^\alpha \exp{\Big(-\frac{L}{L_*}\Big)} \frac{dL}{L_*} \,.
\end{equation}
This is the form given by \citet{schechter76} and is also used for fitting the COPSS II results in \citet{keating+16}.

Other works dealing with Schechter fits of parameterize the function in a number of forms. In their discussion of the effects of cosmic variance on fitted Schechter parameters, \citet{trenti+08} use
\begin{equation}\label{eq:TSpar}
    \phi(L)dL = 
    \phi_{*,{\rm TS08}} \Big(\frac{L}{L_*}\Big)^\alpha \exp{\Big(-\frac{L}{L_*}\Big)} dL \,.
\end{equation}
This differs from equation~\ref{eq:Spar} by the use of $dL$ in place of $dL/L_*$. The result is that the normalization $\phi_{*TS08}$ is a factor of $L*$ smaller than our normalization: $\phi_{*TS08}=\phi_*/L_*$. 

\citet{kelly+08} replace the $\phi_*$ parameter with a normalization depending on the number density of galaxies, $n$\footnote{Technically this work defines the luminosity function in terms of the total number of galaxies in the observable universe, $N$ rather than the number density $n$, and uses $N$ as their parameter. The luminosity function $\phi_{K08}(L)$ presented there is then related to the one given here by $\phi_{K08}(L)=V_{\rm uni}\phi(L)$ and our $N=nV_{\rm uni}$ where $V_{\rm uni}$ is the volume of the observable universe.},
\begin{equation}\label{eq:Kpar}
    \phi(L)dL = 
    \frac{n}{\int \big(\frac{L}{L_*}\big)^\alpha \exp\big(-\frac{L}{L_*}\big) \frac{dL}{L_*}} \Big(\frac{L}{L_*}\Big)^\alpha \exp{\Big(-\frac{L}{L_*}\Big)} \frac{dL}{L_*} \,.
\end{equation}
Comparing Equations~\ref{eq:Spar} and~\ref{eq:Kpar} we see that $n=\phi_* \int \big(\frac{L}{L_*}\big)^\alpha \exp\big(-\frac{L}{L_*}\big) \frac{dL}{L_*}$, which gives rise to our Equation~\ref{eq:schn}.

It is common practice (including in this work) to plot the luminosity function on a logarithmic scale. For such plots the appropriate units are number per unit volume per dex, $\frac{dn}{d\log L}$, rather than linear $\frac{dn}{dL}$. The conversion between these units is given by noting that $dL = L \ln(10) d\log L$, so that
\begin{equation}\label{eq:dexpar}
    \phi_{\rm dex}(L) = L \ln(10) \phi(L) \,.
\end{equation}
The data from COLDz and ASPECS are directly fit in these units. The parameterization for these fits is given by \citet{riechers+18}
\begin{equation}\label{eq:Rpar}
    \phi_{\rm dex}(L) d\log L = 
    \phi_* \Big(\frac{L}{L_*}\Big)^{\alpha_{R19}} \exp \Big(\frac{L}{L_*}\Big) \ln 10 d\log L \,.
\end{equation}
Inserting Equation~\ref{eq:Rpar} into Equation~\ref{eq:dexpar} gives 
\begin{equation}
    \phi(L)dL = 
    \phi_{\rm dex}(L) \frac{1}{L \ln 10} dL = 
    \phi_* \Big(\frac{L}{L_*}\Big)^{\alpha_{R19}-1} \exp \Big(\frac{L}{L_*}\Big) \frac{dL}{L_*} dL \,,
\end{equation}
which shows that the faint-end slope of this parameterization difference from that of Equation~\ref{eq:Spar} with $\alpha_{R19}=\alpha+1$

To summarize, Equations~\ref{eq:Spar} and~\ref{eq:TSpar} both give the number density per linear interval in $L$, and have the same $L_*$ and $\alpha$ parameters, with different, but related $\phi_*$ parameters. Equation~\ref{eq:Rpar} gives the number density per logarithmic interval, which can be converted to the number density per linear interval through division by $L \ln 10$. Equation~\ref{eq:Rpar} has the same value for $L_*$ and $\phi_*$ as Equation~\ref{eq:Spar}, but a value of $\alpha$ that is $1$ greater than the $\alpha$ in either of the other parameterizations. In all cases, the ``normalization'' parameter $\phi_*$ depends on both the number of objects and the shape of the luminosity function. This parameter can be replaced by a mean number density $n$, which separates the normalization and shape, following Equations~\ref{eq:Kpar} and~\ref{eq:schn}.

\section{Schechter Function Fitting}\label{appendix:fitting}

For the Schechter fits reported in this paper, we use the package emcee \citep{foreman-mackey+13} which employs an affine invariant MCMC ensemble sampler to explore the posterior distribution of a set of parameters, $\bm{\Theta}$, given a set of observations, $x$, and a posterior probability function 
\begin{equation}\label{eq:posterior}
    p(\bm{\Theta} | x) \propto 
    p(x | \bm{\Theta}) p(\bm{\Theta}) \,,
\end{equation}
where $p(\bm{\Theta})$ is a prior on the model parameters and $p(x | \bm{\Theta})$ is the likelihood of the observations for a given set of parameters.

We perform our fits using unbinned data, so each observation is the luminosity of an individual object $L_i$. The likelihood of $L_i$ being drawn from a luminosity function with parameters $\bm{\Theta}$ is
\begin{equation}\label{eq:pL}
    p(L_i | \bm{\Theta}) = 
    \frac{\phi(L_i | \bm{\Theta})}{\int \phi(L | \bm{\Theta}) dL} \,.
\end{equation}
It is common to fit luminosity functions by maximizing 
\begin{equation}\label{eq:simpleL}
    p(L|\bm{\Theta}) = 
    \prod_i p(L_i | \bm{\Theta})     \,.
\end{equation}
However, this procedure has no dependence on the normalization of the luminosity function. In other words, maximizing Equation~\ref{eq:simpleL} only determines the shape of the luminosity function. The normalization is usually determined by fixing the integral of the luminosity function times the survey selection function to be equal to the number of observed objects. This procedure does not account for uncertainty in the normalization due to field to field variations in the number of objects.

\citet{kelly+08} correct this by fitting the mean number density of objects of interest $n$ and the shape parameters $\bm{\Theta_S}$ of the luminosity function. For a given number of observed objects $N_{\rm obs}$ The required likelihood is given by
\begin{equation}
    p(L|n,\bm{\Theta}_S) = 
    {nV_{\rm uni} \choose N_{\rm obs}} \times \Big( 1-p({\rm det} | n,\bm{\Theta}_S) \Big)^{nV_{\rm uni}-N_{\rm obs}} \times  \prod_{i=1}^{N_{\rm obs}} \Big( p({\rm det}_i | L_i) p(L_i | n,\bm{\Theta}_S) \Big)  \,.
\end{equation}
Here $V_{\rm uni}$ is the volume of the universe over which the objects of interest can be observed, so that $nV_{\rm uni}$ is the number of objects in the observable universe. $1-p({\rm det} | n,\bm{\Theta}_S)$ is the probability of an object drawn from a luminosity function parameterized by $n$ and $\bm{\Theta}_S$ being undetected in a given survey, which accounts for both the survey sky coverage and sensitivity limits of the observation. The product is taken over all objects observed by the survey, with $p({\rm det}_i | L_i)$ being the probability of source $i$ of luminosity $L_i$ being detected by the survey and $p(L_i | n,\bm{\Theta}_S)$ given by Equation~\ref{eq:pL}.

As fitting $\phi(L|n,\bm{\Theta}_S)$ requires only relative likelihoods for different sets of parameters, we can drop terms independent of $n$ or $\bm{\Theta}$ to get
\begin{equation}\label{eq:pbinom}
    p(L|n,\bm{\Theta}_S) \propto 
    {nV_{\rm uni} \choose N_{\rm obs}} \times \Big( 1-p({\rm det} | n,\bm{\Theta}_S) \Big)^{nV_{\rm uni}} \times  \prod_{i=1}^{N_{\rm obs}} p(L_i | n,\bm{\Theta}_S)  \,.
\end{equation}
In many situations, $nV_{\rm uni}$ is large and the binomial coefficients can become difficult to evaluate. Assuming that the detection probability is small, as is the case for a survey covering only a small part of the sky, we can use the Poisson approximation of the binomial distribution to write the likelihood 
\begin{equation}\label{eq:ppois}
    p(L|n,\bm{\Theta}_S) \propto 
    (nV_{\rm uni})^{N_{\rm obs}} e^{-n\,V_{\rm uni}\,p({\rm det}|n,\bm{\Theta}_S)} \prod_{i=1}^{N_{\rm obs}} p(L_i | n,\bm{\Theta}_S) \,.
\end{equation}
This gives the likelihood we use for Equation~\ref{eq:posterior}.

We parameterize the luminosity function as a Schechter function in the form of Equation~\ref{eq:Kpar}, modified with a lower luminosity cutoff, $L_{\rm min}$, which accounts for the faintest detectable galaxy in our survey:
\begin{equation}\label{eq:KparLmin}
    \phi(L)dL = 
    \frac{n}{\int_{L_{\rm min}}^\infty \big(\frac{L}{L_*}\big)^\alpha \exp\big(-\frac{L}{L_*}\big) \frac{dL}{L_*}} \Big(\frac{L}{L_*}\Big)^\alpha \exp{\Big(-\frac{L}{L_*}\Big)} \frac{dL}{L_*} \,.
\end{equation}
This gives $\bm{\Theta} = (n,L_*,\alpha)$. Our inclusion of the luminosity cutoff has the effect of redefining $n$ as the mean number density of galaxies brighter than $L_{\rm min}$. The likelihood for each object is given by
\begin{equation}
    p(L_i|n,L_*,\alpha) = 
    \frac{\big(\frac{L_i}{L_*}\big)^\alpha \exp{\big(-\frac{L_i}{L_*}\big)}\frac{1}{L_*}}{\Gamma(\alpha+1,L_{\rm min}/L_*)} \,,
\end{equation}
where $\Gamma$ is the upper incomplete gamma function. \footnote{The gamma function is only formally defined when $\alpha+1 > 0$, but as long as $L_{\rm min} > 0$ the integral defining the upper gamma function converges and the probability can be evaluated.}
Our model surveys assume that all galaxies within the survey volume above $L_{\rm min}$ are detected. Therefore the selection function is
\begin{equation}
    {\rm det}_i =
    \begin{dcases}
        1 & L_i > L_{\rm min}, r_i\in V_{\rm obs} \\
        0 & {\rm otherwise}
    \end{dcases}
\end{equation}
where $r_i\in V_{\rm obs}$ indicates the object falls within our survey volume, and
\begin{equation}
    p({\rm det}|\phi_*,L_*,\alpha) = 
    \frac{V_{\rm obs}}{V_{\rm uni}} \,.
\end{equation}
Substituting these into Equation~\ref{eq:ppois} gives
\begin{equation}
    p(L|n,L_*,\alpha) \propto 
    (nV_{\rm uni})^{N_{obs}}
    \exp (-nV_{\rm obs}) \prod_{i=1}^{N_{\rm obs}} \frac{\big(\frac{L_i}{L_*}\big)^\alpha \exp{\big(-\frac{L_i}{L_*}\big)}\frac{1}{L_*}}{\Gamma(\alpha+1,L_{\rm min}/L_*)} \,.
\end{equation}
The priors $p(\bm{\Theta})=p(n,L_*,\alpha)$ could be used to incorporate external constraints on the parameters. Here we are interested in how well individual observations do at constraining the luminosity function, therefore we adopt uninformative priors on $\log n$, $\log L_*$, and $\alpha$.

\section{Redshift Evolution of the Second Moment}\label{appendix:evo2m}

\begin{figure}
    \centering
    \includegraphics{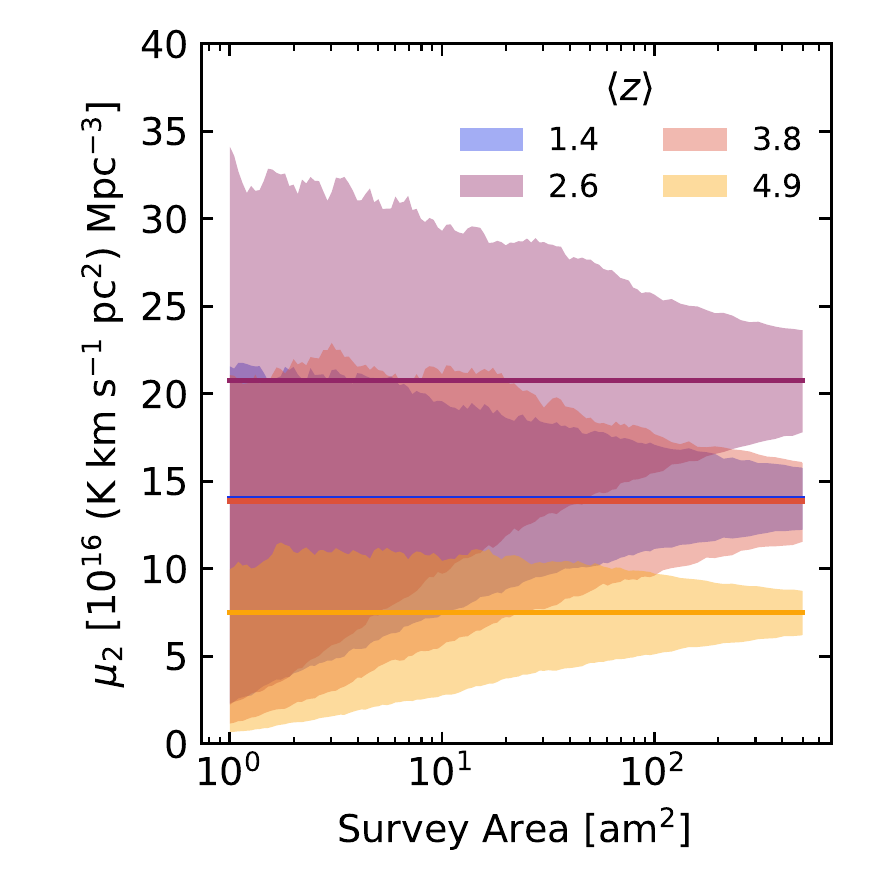}
    \caption{The 16th-84th percentile range as a function of survey area (filled regions) and true value (horizontal line) of the second moment for the same redshifts as Figure~\ref{fig:rs}. Here we show only the combined Poisson and cosmic uncertainties, as the latter contributes minimally to the sample uncertainty in the second moment. Note that we have elected to present results in terms of $\mu_2$ rather than $P_{\rm shot}$ in order to capture only physical changes due to redshift evolution in the luminosity function.}
    \label{fig:rs2m}
\end{figure}

Due to the redshift-dependent conversion between temperature and luminosity units, shot power that is constant with redshift is not equivalent to a constant second moment of the luminosity function. Therefore, to obtain a quantity which shows only evolution in the underlying luminosity function, we convert from shot power units to units of (K km s$^{-1}$ pc$^2$)$^2$ Mpc$^{-3}$ using
\begin{equation}\label{eq:shotmu2}
    \mu_2 = \Big(\frac{H(z)}{(1+z)^2}\Big)^2 P_{\rm shot}.
\end{equation}
In Figure~\ref{fig:rs2m} we show this quantity for our fiducial model and the redshifts considered in Section~\ref{sec:rs}.

\bibliography{refs}

\end{document}